\documentclass[12pt,preprint]{aastex}
\usepackage{epsf}

\def\etal{{et al.}}
\def\eg{{e.g.,}}
\def\ie{{i.e.,\ }}
\def\Htwo{${\rm H_2}~$}

\def\K{~{\rm K}}
\def\kms{~{\rm km~s^{-1}}}
\def\cm3{~{\rm cm^{-3}}}

\def\tcl{t_{\rm cool}}
\def\lcl{l_{\rm cool}}
\def\kpc{~{\rm kpc}}
\def\Msun{~{\rm M}_{\sun}}

\begin{document}

\title{Three-Dimensional Numerical Simulations of
Thermal-Gravitational Instability in Protogalactic Halo Environment} 

\author{Chang Hyun Baek\altaffilmark{1,2}, Hyesung Kang\altaffilmark{1},
Jongsoo Kim\altaffilmark{2} and Dongsu Ryu\altaffilmark{3}}

\altaffiltext{1}{Department of Earth Sciences, Pusan National University, Pusan
            609-735, Korea: \\ chbaek@pusan.ac.kr, kang@uju.es.pusan.ac.kr}
\altaffiltext{2}{Korea Astronomy \& Space Science Institute, Whaam-Dong,
            Yuseong-Gu, Taejeon 305-348, Korea: jskim@kasi.re.kr}
\altaffiltext{3}{Department of Astronomy \& Space Science, Chungnam National
             University, Daejeon 305-764, Korea: \\ ryu@canopus.cnu.ac.kr}

\begin{abstract}
We study thermal-gravitational instability in simplified models
for protogalactic halos
using three-dimensional hydrodynamic simulations.
The simulations started with isothermal density perturbations
of various power spectra, and followed the evolution of gas
with radiative cooling down to $T = 10^4$ K, background heating,
and self-gravity for up to $\sim 20$ cooling times. Then cooled and
condensed clouds were identified and their physical properties were
examined in detail. In our models, the cooling time scale
is several times shorter than the gravitational time scale. Hence, during
early stage clouds start to form around initial density peaks by
thermal instability. Small clouds appear first and they are
pressure-bound. Subsequently, the clouds grow through compression by
the background pressure as well as gravitational infall. During late
stage cloud-cloud collisions become important, and clouds grow mostly
through gravitational merging. Gravitationally bound clouds with mass
$M_c \ga 6 \times 10^6 \Msun$ are found in the late stage. They are
approximately in virial equilibrium and have radius $R_c \simeq 150-200$
pc. Those clouds have gained angular momentum through tidal torque as well
as merging, so they have large angular momentum with the spin parameter
$\left<\lambda_s\right> \sim 0.3$. The clouds formed in a denser background
tend to have smaller spin parameters, since the self-gravity, compared to
the radiative cooling, is relatively less important at higher density.
The \Htwo cooling below $T = 10^4$ K does not drastically change the
evolution and properties of clouds, since it is much less
efficient than the H Ly$\alpha$ cooling. The slope of initial density
power spectrum affects the morphology of cloud distribution, but the
properties of individual clouds do not sensitively depend on it.
We point limitations of our study and mention briefly the implications
of our results on the formation of
protoglobular cluster clouds in protogalactic halos.
\end{abstract}


\section{Introduction}

Thermal instability (TI) is one of key physical processes in
astrophysical environments where optically thin gas cools radiatively
and condenses \citep{fie65}. It has been applied to explain, for instance,
the multiple phases of interstellar gas \citep[\eg][]{fgh69,mo77}, the
formation of globular clusters \citep[\eg][]{fall85}, cooling flows
in clusters of galaxies \citep[\eg][]{nuls86}, and the generation of
turbulent flows in the interstellar medium (ISM) \citep[\eg][]{koy02,kri02}.
Cold dense clumps or clouds that are confined by the background pressure
can form in a hot, radiatively cooling medium via TI 
\citep[\eg\ see][]{burkert02}. In the simplistic picture of TI, those
overdense regions undergo a quasi-static compression in near pressure
equilibrium \citep{fie65}. However, this {\it isobaric} condensation
occurs only when the clouds are small enough to adjust to pressure
changes faster than the gas cools. According to numerical simulations
of the collapse of thermally unstable clouds
\citep[\eg][]{dav88,bri90,mrf90,kang00,bkr03}, the clouds may undergo
a {\it supersonic} compression when the cloud size is comparable to
the cooling scale, $\lcl$ (the distance over which a sound wave travels
in a cooling time), while the clouds much larger than this scale cool
{\it isochorically}.

It has been suggested that the TI could be responsible for the
formation of protoglobular cluster clouds (PGCCs) in protogalactic
halos, which can explain the origin of old halo globular clusters
\citep[\eg][]{fall85,kang00}. 
Among many models of globular cluster formation 
\citep[\eg\ see][]{pjmn99,kravtsov05}, this model based on TI
is classified as a secondary model in which the two-phase of cold
dense clouds in hot background gas was developed through TI in
protogalactic halos and the condensed clouds further collapsed to
become globular clusters. We previously studied the development of TI
in detail using one and two-dimensional numerical simulations with
spherically symmetric and axisymmetric {\it isolated} gas clouds in
static environments of uniform density \citep{kang00, bkr03}.
However, according to the current paradigm of cold dark matter models
of structure formation, large protogalaxies comparable to the Milky Way
formed via hierarchical clustering of smaller systems. Hence, inevitably
density perturbations should exist on a wide range of length scales
inside the protogalactic halos, and {\it an ensemble of} clumps emerge.
In addition, although the thermal process initiates the formation of
the clumps, eventually the self-gravity should become important, since
the gravitational time scale is just a few times longer than the cooling
time scale. As a result the initial clumps grow by both thermal and
gravitational processes to become bound clouds. Hence, the
thermal-gravitational instability should be considered
\citep[\eg\ see][]{balb86}.

In this paper we study the thermal-gravitational instability in
a hot background whose physical parameters are relevant for
protogalactic halo gas.
Three-dimensional simulations were made. The simulations started with
random Gaussian density perturbations of various power spectra. The
evolution of gas under the influence of thermal-gravitational
instability was followed up to the formation of gravitationally bound
clouds. The physical properties of the clouds are examined in detail.
Our goal is to study how self-gravity and gravitational interactions
affect the physical and dynamical properties of clouds that condense
initially via TI. Although our models would be yet too simple to be
directly applied to the real situation, we try to extract the
implications of our results on the formation of PGCCs in protogalactic
halos. In \S 2 our models and numerical
details are described. Simulation results are presented in \S 3,
followed by summary and discussion in \S 4.

\section{Simulations}

\subsection{Models for Protogalactic Halo}

A gas of $T_h = 1.7\times 10^6$K in a cubic, periodic
simulation box was considered. This temperature corresponds to that of
an isothermal sphere with circular velocity $V_c=220\kms$, representing
the halo of disk galaxies like the Milky Way. The fiducial value of
the mean background density of hydrogen nuclei was chosen to be 
$n_h = 0.1 \cm3$. A case with higher density $n_h = 0.3 \cm3$ was also
considered to explore the effects of background density (Model D, see
Table 1 below). For the primordial gas with an assumed ratio of
He/H number densities of 1/10, the gas mass density is given by
$\rho_h = (2.34\times 10^{-24}~{\rm g})~n_h$. With $T_h = 1.7\times 10^6$K
and $n_h = 0.1 \cm3$, the initial cooling time scale is
$\tcl=2\times 10^7$ yrs. On the other hand, the free-fall time scale,
or the gravitational time scale, is $t_{\rm grav} = 1.4 \times 10^8$ yrs,
which is about seven times longer than the cooling time scale. Note that
$\tcl \propto n_h^{-1}$, while $t_{\rm grav} \propto n_h^{-1/2}$.
So cooling, compared to gravitational processes, becomes relatively more
important at higher densities. The cooling
length scale is given as $\lcl = c_h \cdot t_{\rm cool} = 4$ kpc, where
$c_h=198 \kms$ is the sound speed. The simulation box was set to have
the size $L = 10 \kpc = 2.5 \lcl$. It was chosen to be large enough
to accommodate a fair number of thermally unstable clouds of cooling
length size, and so to get fair statistics of cloud properties.

To mimic density perturbations existed on a wide range of length scales
inside the protogalactic halo, the initial density field was drawn from
random Gaussian fluctuations with predefined density power spectrum.
The density power spectrum was assumed to have the following power law
\begin{equation}
P_k  \equiv |\delta \rho_k|^2 4\pi k^2  \propto k^n,
\end{equation}
where $k$ is the three-dimensional wavenumber, 
$k = (k_x^2 + k_y^2 + k_z^2)^{1/2}$.
In order to explore how the initial density perturbations affect the
formation and evolution of clouds, three types of density power spectrum
were considered: white noise with $n=0$ as the representative case, as
well as random fluctuation with $n=2$ (Model R) and Kolmogorov spectrum
with $n=-5/3$ (Model K). Only the powers with $k$'s corresponding to
wavelengths $\lambda \leq L/2$ were included. The amplitude of
density power spectrum was fixed by the condition $\delta_{\rm rms}
\equiv \left<\delta \rho^2 \right>^{1/2}/\left<\rho\right> = 0.2$.
The initial temperature was set to be uniform, assuming {\it isothermal}
density perturbations. The initial velocity was set to be zero everywhere
in the simulation box.

\subsection{Numerical Details}

The gas-dynamical equations in the Cartesian coordinate system including
self-gravity, radiative cooling and heating are written as
\begin{equation}
{\partial \rho \over \partial t} + \nabla \cdot (\rho \mathbf v) = 0,
\end{equation}
\begin{equation}
{\partial \over \partial t} (\rho \mathbf v) 
+ \nabla \cdot (\rho \mathbf v \mathbf v + p \mathbf I) = -\rho \nabla \Phi,
\end{equation}
\begin{equation}
{\partial E \over \partial t} + \nabla \cdot [ (E + p) \mathbf v] =
 -\rho \mathbf v \cdot \nabla \Phi + \Gamma - \Lambda,
\end{equation}
\begin{equation}
\nabla ^2 \Phi = 4 \pi G \rho,
\end{equation}
where $E = (1/2) \rho v^2 + p/(\gamma-1)$, $\Lambda$ and $\Gamma$ are
the cooling and heating rates per unit volume, and the rest of the
variables have their usual meanings. For the adiabatic index, $\gamma=5/3$
was assumed.

The hydrodynamic part was solved using an Eulerian hydrodynamics code
based  on the total variation diminishing (TVD) scheme \citep{ryu93}.
The version parallelized with the Message-Passing Interface (MPI) library
was used.

The self-gravity, cooling and heating were treated after the hydrodynamic
step. For the self-gravity, the gravitational potential was calculated by
the usual Fast Fourier Transform (FFT) technique. Then, the gravitational
force was implemented in a way to ensure second-order accuracy as
\begin{equation}
\mathbf v^{n+1} = \mathbf v^{n+1,*}
- \Delta t^n \cdot \nabla\Phi^{n+1/2},
\end{equation}
where $\mathbf v^{n+1,*}$ is the velocity updated in the hydrodynamic
step and $\Phi^{n+1/2}$ is the potential calculated with
$\rho^{n+1/2} \equiv (\rho^n + \rho^{n+1})/2$.

For the radiative cooling, the collisional ionization equilibrium (CIE)
cooling rate for a zero-metalicity (primordial), optically thin gas in
the temperature range of $10^4$K $\le T \le$ $10^{7.5}$K was adopted 
\citep{suth93}. The cooling function, $L(T) \equiv \Lambda/n_H^2$, is
plotted in Figure 1 as the solid line. With $L(T) = 0$ for $T < 10^4$K,
it was assumed that the extra-galactic/stellar UV radiation
photo-dissociates \Htwo molecules in the protogalactic halo and so
prohibits the gas from cooling below $10^4$ K. Accordingly, the minimum
temperature was set to be $T_{\rm min}=10^4$K. Note that the CIE cooling
rate is higher than the cooling rate based on the non-equilibrium
ionization (NEQ), which had been adopted in our previous two-dimensional
simulations, especially near the H and He $\rm {Ly\alpha}$ line emission
peaks \citep[see Figure 1 of][]{bkr03}. The higher cooling rate was
chosen to accentuate the effects of cooling over those of
gravitational processes.

If \Htwo molecules have formed efficiently via gas phase reactions enough
to be self-shielded from the photo-dissociating UV radiation, or if the
halo gas had been enriched by metals from first-generation supernovae,
however, the gas would have cooled well below $10^4$ K. In order to
explore how the additional cooling below $10^4$K affects the formation
and evolution of clouds, in a comparison model (Model C), the following
{\it mock} cooling function in the range of $10^2$K $\le T \le $ $10^4$K 
was adopted
\begin{equation}
L_{\rm H_2} = 1.0 \times 10^{-26} \exp\left(-{10^3 \over T}\right)
{\rm erg~cm^3~s^{-1}}.
\end{equation}
Although this mock cooling rate was designed to represent typical \Htwo
ro-vibrational line emissions for the gas with \Htwo abundance of
$n_{\rm H_2}/n_H \sim 10^{-3}$ \citep{shkang87}, the exact amplitude and 
form of $L_{\rm H_2}$ are not important in our discussion. Figure 1 shows
$L_{\rm H_2}$ as the dashed line. The minimum temperature  was set to be
$T_{\rm min} = 10^2$K in the C model.

It was assumed that the background gas was
initially under thermal balance and there existed a constant background heating
equal to the cooling of the initial background gas, that is,
\begin{equation}
\Gamma = L(T_h) n_h ^2.
\end{equation}
To prevent any spurious heating the highest temperature was set to be
$T_{\rm max}=T_h$. This ad-hoc heating was applied in order to maintain
the temperature of the background gas with initial mean density at
$T \simeq T_h$. This can be provided by several physical processes such as
turbulence, shocks, stellar winds and supernova explosions in protogalaxies.
Without this heating the background gas would have cooled down in a
few $\tcl$.

It has been pointed out that the thermal conduction can affect profoundly
TI in the ISM \citep[\eg][]{mb90,koy04}. With the Spitzer value of
thermal conductivity for ionized gas,
$\kappa \sim 5 \times 10^{-7}T^{5/2} {\rm erg~s^{-1}cm^{-1}K^{-1}}$
\citep{spitz79}, the thermal conduction time scale is
\begin{equation}
t_{\rm cond} \sim 2 \times 10^9 \left(\frac{T}{1.7 \times 10^6 {\rm K}}
\right)^{-5/2} \left(\frac{n}{10^{-1}{\rm cm^{-3}}}\right) \left(\frac{L}
{1~{\rm kpc}}\right)^2 {\rm yrs}.
\end{equation}
It shows that $t_{\rm cond}$ is much longer than $t_{\rm cool}$ 
across the cooling length ($l_{\rm cool} = 4$ kpc),
so the thermal
conduction is expected to be negligible. 
But the clouds finally emerged in our simulations have radius
of $R_c \sim 150 - 200$ pc (see \S 3.3). Even in such scale $t_{\rm cond}$ is still
a few times longer than, or at most comparable to, $t_{\rm cool}$.
Also those clouds are gravitationally bound, so the thermal evaporation 
should not be important. In addition, weak magnetic field, if it
exists, may reduce significantly the value of thermal conductivity from
the Spitzer value \citep[\eg][]{cc98}, although we do not explicitly
include magnetic field in this work. All together, it is expected that
the thermal conduction does not play a major role in the regime we are
interested in, and hence we ignored it in our simulations.

\subsection{Model Parameters}

Simulations were made with $1024^3$, $512^3$ and $256^3$ grid zones,
allowing a uniform spatial resolution of $\Delta l = 9.8 - 39$ pc.
Simulations started at $t = 0$ and lasted up to $t_{\rm end} =16 - 20 \tcl$.
This terminal time corresponds to $\sim 2 -3 t_{\rm grav}$, so
gravitational bound clouds should have emerged by then.

Total seven simulations are presented in this paper, which differ in numerical
resolution, the power spectrum of initial density perturbations, cooling,
and mean background density. Model parameters are summarized in Table 1.

\section{Results}

\subsection{Evolution of Halo Gas}

We start to describe the results by looking at the global
evolution of gas and its morphological distribution. Figure 2
shows the density power spectrum at different times in the S1024 model
where $P_k \propto {\rm constant}$ initially. In the figure the
dimensionless wavenumber is given as $k \equiv L/\lambda$, which counts
the number of waves with wavelength $\lambda$ inside the size $L$ of
simulation box. The power spectrum is presented in a way that
\begin{equation}
\int P_k dk = \left< \rho^2 \right>.
\end{equation}

With $t_{\rm cool} < t_{\rm grav}$, initially TI should work. The
noticeable features in the early evolution of power spectrum are
the followings: during $t \la 2\tcl$ the powers with
$k \ga 40$ (or $\lambda \la \lcl/16$) are reduced significantly, and
then the powers on those small scales grow back by $\sim 4\tcl$. The
initial decrease of small scale powers is a consequence of initial
{\it isothermal} density perturbations. The accompanying pressure
fluctuations have generated sound waves, and those sound waves have ironed
out the perturbations of small scales. The follow-up, fast growth of
small scale powers is due to {\it nonlinear} behavior of TI.
Although the {\it linear} growth rate is independent of scale (for
$\lambda < l_{\rm cool}$), the growth can be limited once the density
increases and the cooling length becomes smaller than the perturbation
scale. With the perturbation scale smaller than the cooling length, the
further condensation progresses {\it isochorically} and the growth slows
down. As a result, small scale clumps appear first. This point has been
made previously, for example, by \citet{burkert02}. In addition, when
perturbations get compressed by the background pressure, the density in
the central region increases first and clouds form inside out. This
contributes the fast growth of small scale powers too. By the end of
this early TI stage, $\sim 4\tcl$, the power spectrum peaks at
$\lambda \sim \lcl/50$. 

After $\sim 4\tcl$, the self-gravity starts to play a role. Clouds
grow through gravitational infall as well as compression by the
background pressure. Eventually, massive clouds form through cloud-cloud 
collision, or merging among clouds (see below). During these stages,
the powers grow over all scale. At the same time the peak shifts to
smaller wavenumbers, reflecting the appearance of larger, massive clouds

We note that with periodic boundary, once the power of the scale
corresponding to the box size reaches nonlinear, the large scale
clustering becomes saturated. The power spectrum in Figure 2 shows that
in the S1024 model the scale which has gone nonlinear by the end is
$\sim L/2 - L/3$, indicating that the assumption of periodic boundary
should not have affected our results significantly. In any case
our major focus lies on the properties of individual clouds rather
than their clustering.

In order to show how clouds form and grow as well as how their
distribution evolves, three-dimensional iso-density surfaces at four
different times in the S1024 model are presented in Figure 3. As
noted above, at the end of the early TI stage ($\sim 4 \tcl$)
mostly small clouds appear. By the end of the follow-up stage
of TI and gravitational growth, $\sim 8 \tcl$, the clouds
become larger. During late stage, the clouds become even larger
but their number reduces, as a result of cloud-cloud mergers,
which can be seen in the bottom two images.

\subsection{Identification of Clouds and Their Mass Function}

In order to study the properties of formed clouds they were identified
using the algorithm CLUMPFIND described by \citet{Will94}. The algorithm
basically tags cells around a density peak as the ``cloud cells'', if they
satisfy the prescribed criteria of density and temperature. We chose
the following criteria: $\rho \ge 10 \rho_0$ and $T \leq 10^5 $K. Here
$\rho_0$ is the {\it initial} mean density. There is an arbitrariness in
these threshold values. But the identification of clouds does not
depend sensitively on the choices of threshold values, since clouds are
well delineated by rather sharp jumps in density and temperature. In
addition, we qualified only those with at least $3 \times 3 \times 3$
cells or more as clouds. Once clouds were identified, their various
quantities were calculated.

The first row of Figure 4 shows the number of clouds, $N_c$, as a
function of the cloud mass, $M_c$, at the times same as those in Figure 3
in the S1024 model. By $4 \tcl$ a significant number of clouds,
$N_c \sim 3 \times 10^3$, form through the growth of initial high
density peaks by TI. The mass function at $4 \tcl$ is roughly
{\it Gaussian}, since the initial density perturbations were drown from
a random Gaussian distribution. During the follow-up stage of TI and
gravitational growth, more peaks develop into clouds and at the same
time they become massive. The mass function evolves roughly into the
{\it log-normal} distribution, shown at $8 \tcl$. The log-normal
distribution is a signature of nonlinear structure formation, as reported
in various simulations \citep[see, \eg][]{vaz94,wn01}. During late
stage, more massive clouds develop through gravitational merging
as pointed with Figure 3, and so the mass function extends to higher
mass. The high mass tail of the mass function beyond the peak follows
approximately a {\it power-law} distribution. When the mass function
is fitted to $dN/dM_c \propto M_c^{-\alpha}$ for $M_c > 10^{5.5} \Msun$,
the value of $\alpha$ decreases from $\sim 0.8$ at $12 \tcl$ to
$\sim 0.6$ at $16 \tcl$.

\subsection{Size, Density and Energetics of Clouds}

With the mass, $M_c$, and volume, $V_c$, of identified clouds, the
effective radius was taken as $R_c \equiv (3 V_c /4 \pi)^{1/3}$ and
the mean density was calculated as $\left<\rho_c\right> = M_c/V_c$.
The second and third rows of Figure 4 show $R_c$ and
$\left<\rho_c\right>$ in the S1024 model. For energetics the kinetic
energy relative to the cloud's center of mass, $E_K$, the thermal energy,
$E_T$, and the gravitational energy, $E_G$, were calculated for
identified clouds. The procedure by which the gravitational energy was
calculated is presented in Appendix A along with a note of caution.
The fourth row of Figure 4 shows the ratio of positive to negative energies,
or the {\it virial} parameter
\begin{equation}
\beta \equiv {{2 \cdot (E_K+E_T)} \over |E_G|}.
\end{equation}
The parameter
$\beta$ tells whether clouds are primarily pressure-bound ($\beta \ga 2$)
or gravitationally bound ($\beta \la 2$). Among the gravitationally bound
clouds, the condition $\beta \sim 1$ indicates that they are approximately
in virial equilibrium, although, strictly speaking, the condition applies
only for a stable system where the external pressure is negligible and
the moment of inertia does not change with time.

During the early stage of TI ($\la 4 \tcl$), as small clumps grow
isobarically, their density increases gradually, but
$\left<\rho_c\right> / \rho_0$ has not yet reached the isobaric factor of
$\sim (T_h/T_{\rm min}) \simeq 100$. The self-gravity is negligible with
$\beta \gg 2$ during this stage.
In the follow-up stage of TI and gravitational growth ($\la 8 \tcl$),
pressure-bound clouds develop fully. Those clouds have roughly
$\left<\rho_c\right> / \rho_0 \sim 100$, the maximum isobaric increase
from the initial density, and the mean density is a bit higher in clouds
with larger mass. Their radius scales roughly as $R_c \sim M_c^{0.3}$,
which can be seen in the second row of Figure 4. The virial parameter has
$\beta > 2$ for all clouds, confirming that they are still gravitationally
unbound. The pressure-bound clouds follow  $\beta \propto M_c^{-1}$, which
can be understood as follows. In those clouds, the thermal energy is
dominant over the kinetic energy, \ie  $E_T \sim {\rm a~few} \times E_K$.
With $T \sim T_{\rm min}$ in those clouds, the thermal energy scales as
$E_T \propto M_c$. On the other hand, the gravitational energy scales
approximately as $E_G \propto M_c^2$, since larger clouds have slightly
more concentrated mass distribution.

As clouds grow further primarily through merging in late stage, some
of them become massive enough to be gravitationally bound. After that point
clouds can be divided into two populations of distinct properties
(see the last two times of Figure 4): pressure-bound
clouds with smaller mass and gravitationally bound clouds with larger
mass. The pressure-bound clouds
have the properties similar to those found in the earlier stage. They
have mean density $\left<\rho_c\right> \sim 100 \left<\rho_{bg}\right>$,
where $\left<\rho_{bg}\right>$ is the mean background density. Note that
the background density continues to decrease as more mass goes to clouds.
By the end of the S1024 simulation, $16 \tcl$, only $\sim 1/4$ of gas
mass remains in the background. So the mean density of the pressure-bound
clouds decreases with time. Those pressure-bound clouds have $\beta > 2$
and follow $\beta \propto M_c^{-1}$. 
The gravitationally bound clouds appear first at $\sim 9 \tcl$ or
$\sim 1.3 t_{\rm grav}$ in the S1024 model. In those clouds
the self-gravity enhances the density and the mean density of the
clouds reaches up to $10^4 \left<\rho_{bg}\right>$ or even higher by
$16 \tcl$. The fourth raw of Figure 4 shows that the gravitationally
bound clouds are approximately in virial equilibrium with $\beta \sim 1$.
In addition, we found that in those clouds the kinetic energy is not
smaller but sometimes larger than the thermal energy as expected in a
virialized system, and both the gravitational energy, $E_G$, and the
positive energy, $E_K + E_T$, scale as $M_c^2$.

Two points are noted on the gravitationally bound clouds.
1) Although these clouds have $\beta \sim 1$, they are not in steady-state.
The clouds lose the positive energy rather quickly through cooling and
contract further. But at the same time the cloud mass continues to grow
through merging.
2) The gravitational energy of these gravitationally bound clouds scales as
$E_G \propto M_c^2$, because their radius is in a relatively narrow range of
$150-200$ pc regardless of their mass, as shown in the second row of Figure
4. Since there is no physically obvious reason why the clouds with different 
mass should have similar radii, it should be understood as the result of 
dynamical evolution. However,
their radius is somewhat larger at $16 \tcl$ than at $12 \tcl$. This is
related to the increase of angular momentum in the clouds with time
(see the next subsection).

The distinction between the pressure-bound and gravitationally bound
clouds in late stage can be understood with the {\it critical} mass
\begin{equation}
M_{\rm crit} = 1.18 \left(k T_c \over \mu m_H\right)^2 G^{-3/2} p_{bg}^{-1/2},
\end{equation}
which is the maximum stable mass for an isothermal sphere confined by the
background pressure $p_{bg}$ \citep[see {\it e.g.,}][]{mcc57,kang00}. For
$T_c = 10^4$ K and the background pressure four times smaller than the
initial pressure (due to decrease in the background density), the critical
mass is $M_{\rm crit} \simeq 6 \times 10^6 \Msun$, which coincides well
with the transition mass scale in the last two times of Figure 4.
Incidentally, this mass is similar to the Jeans mass of clouds, which
is given as
\begin{equation}
M_J \equiv \rho \lambda_J^3 = 6 \times 10^6 \left(T_c \over 10^4 \K
\right)^{3/2} \left(n_c \over 100 \cm3\right)^{-1/2},
\end{equation}
where $\lambda_J$ is the Jeans length \citep[see {\it e.g.,}][]{spitz79}.

Although the simulated model would be too simple to represent a real
protogalactic halo,
it is tempting to regard the gravitationally bound clouds as possible
candidates for PGCCs.
That is, some of them may further
cool down below $10^4 \K$ either by UV self-shielding of H$_2$ molecules
or by self-enrichment of metals due to first generation Type
II supernovae, possibly leading to star formation. If about ten of them
turn into globular clusters, their number density would be
$\sim$ 0.01 clusters/kpc$^3$. However, the typical size of globular
clusters $\sim 10$ pc or so. So the clouds should contract further by more
than a factor of 10. But the further collapse is controlled by the rotation
of clouds (see \S 3.4 below). In addition,
the typical mass of globular clusters is $\sim 10^6 \Msun$ or so. So
if these gravitationally bound clouds were to become globular clusters,
the star formation efficiency should be $\sim 10\%$ or so with
$\sim 90 \%$ of their mass dispersed back to protogalactic halo.

It was pointed out by \citet{tkmh97} that ``artificial fragmentation'' due
to errors arising from discretization occurs in numerical simulations
with self-gravity. They argued that the artificial fragmentation can be
suppressed if resolution is maintained high enough that, for instance,
the ``Jeans number'' $\Delta l / \lambda_J \la 0.25$ or so for isothermal
collapses. Here $\lambda_J$ is the local Jeans length. We found that
although the constraint was not complied in a few high density zones,
$\Delta l / \lambda_J$ was kept to be always smaller than 0.4 in the
S1024 simulation. In any case, we followed only up to the formation of
bound clouds, not the subsequent evolution leading to fragmentation of
those clouds. So no obvious fragmentation was observed.

\subsection{Rotation of Clouds}

An important property of clouds that controls the dynamical state and
affects the eventual fate is their rotation. To quantify it, the angular
momentum of clouds relative to their center of mass, $J_c$, was calculated.
The bottom row of Figure 4 shows the specific angular momentum of clouds,
$j_c = J_c/M_c$, in the S1024 model.

In pressure-bound clouds, rotation plays a minor role in their dynamical
evolution, since the rotational energy is smaller by an order of magnitude
than the thermal energy.
But we found that rotation can be dynamically important in
gravitationally bound clouds. The specific angular momentum of those
gravitationally bound clouds is larger than that of pressure-bound clouds,
which can be seen clearly at $16 \tcl$. In fact the gravitationally bound
clouds with a same $M_c$ have higher angular momentum at later time.
For example, the clouds with $M_c \sim 10^{7.5} \Msun$ have a few times
larger $j_c$ at $16 \tcl$ than at $12 \tcl$. This is because the clouds
have grown mostly through merging in late stage and they have gained
angular momentum through merging as well as tidal torque. Its direct
consequence is that the clouds at later stage have larger radius, as
noted in the previous subsection. We note that the clouds in our
simulations are ever evolving. So it is not meaningful to define the
canonical properties of clouds such as $R_c$ and $j_c$ as a function of
$M_c$.

The rotation of gravitationally bound objects is often characterized
by the dimensionless {\it spin} parameter \citep{peebles69}
\begin{equation}
\lambda_s = {{J_c|E_G^{1/2}|} \over {GM_c^{5/2}}}.
\end{equation}
It measures the degree of rational support of the systems with
\begin{equation}
{f_{\rm centrifugal} \over f_{\rm gravity}} \sim {J_c \over R_c^3}
{R_c^2 \over GM_c} \sim \lambda_s^2.
\end{equation}
In the context of cosmological $N$-body simulations, it has been shown that
the typical value of $\lambda_s$ for gravitationally bound objects which
acquired angular momentum through gravitational torque is $\sim 0.05$
\citep[\eg\ see][]{be87}. In Figure 5 red squares represent the spin
parameter for the gravitationally bound clouds with $M_c \geq 10^7 \Msun$
at $16 \tcl$ in the S1024 model. It is clear that $\lambda_s$ of
our gravitationally bound clouds is significantly larger than 0.05. All
except one have $\lambda_s > 0.05$ and the mean value is
$\left<\lambda_s\right> \sim 0.3$. It is because the clouds have gained
angular momentum through merging as well as torque. 

With such large values of $\lambda_s$, the degree of rotational support
of the gravitationally bound clouds should be already substantial.
Hence, rotation should be the key parameter that determines whether some
of those clouds could become PGCCs and collapse further to globular clusters.
For instance, in a dissipative collapse that conserves angular momentum,
the clouds with $\lambda_s \sim 0.3$ can contract only by a factor of a
few, before they become completely rotationally supported and disk-shaped.
However, yet it is not clear whether we should conclude that those clouds
with $\lambda_s \sim 0.3$ can not evolve into globular
clusters. It is because we can not rule out the possibility that the clouds
may be able to lose most, say $\sim 99 \%$, of their angular momentum,
while they collapse and lose $\sim 90\%$ of their mass. For instance,
in the so-called {\it self-enrichment model} of globular cluster formation,
first generation Type II supernovae govern star formation and the removal
of residual gas \citep[\eg\ see][]{pjmn99,sw87}. In this model the star
formation may have occurred preferentially at the core, and most gas in
the outskirt with large angular momentum may have been blown out.

In addition, as noted above, the gravitationally bound clouds found at an
earlier time have smaller angular momentum. So if the clouds were detached
from background and started to collapse earlier, the angular momentum
restriction would be somewhat less severe. Also the clouds emerged from
different environments could have smaller angular momentum (for instance,
see \S 3.8). In any case, subsequent evolution of the gravitationally
bound clouds is beyond the scope of this paper, so the possible connection
between these clouds and PGCCs should be left as a future study.

\subsection{Shape of Clouds}

Shape is another property of clouds that reflects their dynamical state.
In order to quantify it, we examined the {\it shape} parameters defined as
\begin{equation}
q \equiv {b \over a} \qquad {\rm and}  \qquad s \equiv {c \over a},
\end{equation}
with each cloud fitted to triaxial ellipsoid with axes of
$a \geq b \geq c$. The shape parameters have been commonly used to study
clumps in numerical simulations \citep[\eg][]{curir93,gam03}. In order to
find them, first the moment of inertia tensor,
\begin{equation}
I_{ij} = \int \rho x_i x_j d^3{\vec x},
\end{equation}
was constructed for each cloud. Here $\vec x$ is the displacement relative
to the cloud's center of mass and the integral is taken over the cloud
volume. Then, from the three eigenvalues of the tensor,
$I_{11} \geq I_{22} \geq I_{33}$, the shape parameters were calculated as 
\begin{equation}
q = \left( { I_{22} \over I_{11} } \right)^{1/2} \qquad {\rm and}  \qquad
s = \left( { I_{33} \over I_{11} } \right)^{1/2}.
\end{equation}
The clouds with $s \sim q < 1$ are of prolate shape and the clouds with
$s < 1$ and $q \sim 1$ are of oblate shape, while triaxial clouds
have $s < q < 1$. Spherical clouds have $s \sim q \sim 1$.

In the previous study of isolated, thermally unstable clouds  using
two-dimensional simulations in cylindrical geometry, we showed that the
cloud shape changes in the course of the evolution \citep{bkr03}. The
degree of oblateness or prolateness is enhanced during the initial
cooling phase, as expected. But it can be reversed later due to the
supersonic infall along the direction perpendicular to the initial
flatness or elongation.

In Figure 6 dots represent clouds in the $q-s$ plane at four times in
the S1024 model. Lines divide the domain into three regions of roughly
prolate (left), triaxial (middle) and oblate (right) shapes. The panel
for $4 \tcl$ indicates that most clouds formed as a result of TI at
early stage are preferentially of prolate shape. This is can be understood
from the fact that filaments are the morphology dominant next to knots of
clouds. By $8 \tcl$ some of prolate clouds have been transformed to
be oblate, as the result of the shape reversal, which was observed in the
two-dimensional study. However, the figure shows that in late stage clouds
tend to shift back to be prolate again. This is because gravitational
merging results preferentially in clouds of elongated prolate shape.

The gravitationally bound clouds with $M_c \geq 10^7 \Msun$, shown in
Figure 5, are marked with red circles in the panel for $16 \tcl$. It is
interesting to note that unlike most clouds, those massive clouds are
preferentially of oblate shape. It is because 
those clouds have large angular momentum and a substantial
degree of rational support.

\subsection{Numerical Convergence}

Convergence is an important issue in any numerical simulations. We tested
it by comparing the results of simulations with different resolution,
\ie S1024, S0512 and S0256 models (see Table 1). Figure 7 shows the
density power spectrum of the S0512 and S0256 models, which can compared
with that of the S1024 model in Figure 2. Note that in our simulations
the amplitude of the initial power spectrum was set by the condition
\begin{equation}
\int_{\rm all} P_k dk \equiv A \int_{\rm all} k^n dk
= (1 + \delta_{\rm rms}^2) \left< \rho \right>^2
\end{equation}
with $\delta_{\rm rms} = 0.2$ for all resolutions. Simulations with
different resolution covers different range of wavenumbers. 
So the simulations of lower resolution started
with larger amplitude, as shown in the figure. Otherwise, the evolution
of the power spectrum looks similar in all three models. For instance,
the two most noticeable features in Figure 2, \ie the initial decrease and
follow-up fast growth of small scale powers, are also present in Figure 7.
But an interesting point is that the scale that suffered the initial
decrease is insensitive to resolution, because it was caused by smoothing
due traveling sound waves. On the other hand, the scale of the peak in
the power spectrum after the follow-up fast growth does depend on
resolution. As a matter of fact, the peak at $4\tcl$ in the S0256 model
occurs at the scale almost four times larger than in the S1024 model.
This indicates that the small scale growth in our simulations was limited
by resolution, as expected.

Figure 8 shows the mass function, virial parameter and specific
angular momentum of clouds in the three models. 
The left panels show the results at early TI stage, while the right panels
show the results at late merging stage.
For early stage a different time was chosen for each model so that
the density power spectrum has a similar amplitude.
In late stage, however, the differences caused by the initial amplitude of
power spectrum become insignificant, so $15~\tcl$ is chosen
for all three models. 
We see that the mass function have been
converged for massive clouds, \ie those with $M_c \ga 10^6 \Msun$ at the
early stage and those with $M_c \ga 10^7 \Msun$ at $15~\tcl$. However, the
number of smaller mass clouds depends on numerical resolution, as expected.
With the minimum number of $3^3$ zones for identified clouds, the
minimum mass scales $M_{\rm min} \propto (\Delta l)^3$. On the contrary,
at the early stage the virial parameter is larger in lower resolutions for
clouds of all mass. It is because the clouds are less compact in lower
resolution, and so their gravitational energy is smaller. But as the clouds
grow more massive and larger in late stage, the difference in the virial
parameter becomes smaller. The angular momentum of clouds is somewhat
smaller in lower resolution at the early stage. It is partly because the
time of the plot is different in different models. However, in late stage
the angular momentum becomes comparable in all three models.

Overall, the formation of small mass clouds, initially through TI and
subsequently by compression and infall,
was affected by resolution in our simulations.
However, we found that the massive, gravitationally bound clouds, which
have formed mostly through gravitational merging, have the properties
which are almost converged.

\subsection{Effects of Initial Power Spectrum}

The effects of different initial perturbations on the formation and evolution
of clouds were examined by comparing the results of simulations with
different initial density power spectrum, \ie the K0512 and R0512 models,
to those of the S0512 model (see Table 1). The K0512 model started with
more power on larger scales, while the R0512 model started with more power
on smaller scales. Figure 9 shows the density power spectrum of the K0512
and R0512 models. In both models, small scale powers suffered the initial
decrease, as in the S0512 model. Especially most of the powers in
$k \ga 40$ were erased substantially in the R0512 model. Hence, the overall
growth was delayed in the R0512 model. On the other hand, the growth
proceeded faster in the K0512 model, with more powers on large scales in 
the beginning. 

Figure 10 shows the total number of clouds as a function of time in the
three models with  different initial density power spectrum. The overall
evolution is similar; the number of clouds increases during the TI and
follow-up growth stages, but eventually decreases as gravitational merging
progresses. But as noted above, the K0512 model evolves first and the S0512
and R0512 models follow. So clouds form from $t \sim 2\tcl$ in K0512, from
$t \sim 5\tcl$ in S0512 and from $t \sim 10\tcl$ in R0512. However, an
interesting point is that the maximum number of clouds is about the same
with $N_{\max} \sim 3000$ in all three models.

In Figure 11 three-dimensional iso-density surfaces are plotted at two sets
of times in the three models. Different times were chosen in different
models, since the formation and evolution of clouds proceeds differently.
The upper panels show the surfaces when the number of clouds is
highest, \ie at $3\tcl$ for K0512, $6\tcl$ for S0512 and 
$11\tcl$ for R0512. The lower panels show the surfaces after
the number of clouds have decreased a little bit due to merging, \ie at
$4\tcl$ for K0512, at $10\tcl$ for S0512 and at $19\tcl$ for
R0512. The most noticeable feature is that the distribution is more
``filamentary'' in the K0512 model, but less ``filamentary'' in the R0512
model, than in the S0512 model. It is because that the initial large scale
powers, which were largest in the K0512 model but almost absent in the
R0512 model, have been developed into significant coherent structures in
the cloud distribution.

Figure 12 shows the mass function, virial parameter and specific angular
momentum of clouds in the three models. The same two sets of times as those
in Figure 11 were chosen. Clouds in the K0512 (R0512) model are
slightly more (less) massive in the left panels and slightly less (more)
massive in the right panels. However, considering the difference in the
plotted time, the cloud mass function should be regarded as reasonably
similar in the three models. On the other hand, the virial parameter of
pressure-bound clouds follows the same diagonal strip in all three
models. There is a spread in the distribution of angular momentum, again
partly because the plotted time is different in different models. But
the angular momentum of the clouds in high mass tail is similar. So we
conclude that the properties of individual clouds, especially for massive
clouds, are not sensitive to the initial perturbations, while the spatial
distribution of clouds reflects the slope of initial density power spectrum.

\subsection{Effects of Different Density and Cooling}

The effects of gas density on the formation and evolution of clouds
were examined with the D0512 model, which has the background density 3
times larger than that of the S0512 mode. Otherwise the two models
are identical (see Table 1). With $\tcl/t_{\rm grav} \propto n_h^{-1/2}$,
the gravity is relatively less important in the D0512 model.
Figure 13 compares the mass function, virial parameter and specific
angular momentum of clouds in the D0512 model (red lines and dotes) with
those in the S0512 model at an early stage of TI ($5\tcl$) and at
a late stage of merging ($15\tcl$), respectively. In the D0512
model there are more clouds with larger mass, reflecting the higher
background density. Yet, the virial parameter is almost identical in
the two models, indicating that the background density is not important
in determining the energetics of individual clouds. As a matter of fact,
we found that the properties of clouds, except the angular momentum,
are not sensitive to the background density. The angular momentum of the
clouds identified in the early stage of TI is similar in the two models.
But in late stage the angular momentum is noticeably smaller in the
higher background density model. The same trend is also obvious in
the spin parameter of gravitationally bound clouds, which is shown for
$M_c \geq 10^7 \Msun$ in Figure 5. While the spin parameter for the S0512
model (blue triangles) does not differ much from that for the higher
resolution S1024 model (red squares), the spin parameter for the D0512
model (green circles) is substantially smaller with the median value of
$\lambda_{\rm med} \sim 0.12$. This is because the clouds formed in
higher background density have experienced relatively less tidal torque
and gravitational merging, through which they have acquired angular
momentum.

The effects of possible cooling further below $10^4 \K$ were examined
with the C0512 model, which includes a {\it mock} \Htwo cooling given
in equation (7). Otherwise it is same as the S0512 model
(see Table 1). Figure 13 compares the mass function, virial parameter and
specific angular momentum of clouds in the C0512 model (black lines and
dotes) with those in the S0512 model. The cloud properties are similar
overall, except the smaller thermal energy in the C0512 model, which is
expected from the additional cooling. Because the thermal energy counts
for most of the positive energy, especially in pressure-bound clouds,
the virial parameter is smaller in the C0512 model. But in massive,
gravitational bound clouds, the kinetic energy is comparable to or
sometimes larger than the thermal energy, as noted in \S 3.3. So the
effects of the additional cooling is less important in those clouds.

To quantify how much the additional cooling changes the thermal state of
gas, we compare the mass distribution, $f(T)$, for the S0512 and C0512
models in Figure 14. The gas with $T \sim 10^4 \K$ in S0512 spreads below
$10^4 \K$ in C0512, as expected. In addition some of the gas with
$10^4 \la T \la 2 \times 10^4 \K$ has cooled below $10^4 \K$. But the
additional cooling does not affect much the gas of higher temperature.
Even with this additional cooling the mass fraction peaks still at
$T \sim 1.5 \times 10^4 \K$ and most of the ``cloud gas'' has
$10^3 \la T \la 10^5 \K$, not only because the assumed mock \Htwo
cooling is inefficient, but also  because some of the gas has been
reheated by shocks and compression. 

Here we should note that with lower temperature, the
clouds in the C0512 model have smaller Jeans mass and could go through
further fragmentation. But with a fixed grid resolution, the simulation
could not follow it. In fact, in the C0512 model, the Jeans number,
$\Delta l/\lambda_J$, reached up to $\ga 1$ in the center of some
gravitationally bound clouds. However, as noted above, we did not intend
to follow such fragmentation in this study.

\section{Summary and Discussion}

We study the role of self-gravity and gravitational interactions 
in the formation of clouds via thermal instability (TI) 
through three-dimensional hydrodynamic simulations.
We considered the gas in protogalactic halo environment
with $T \sim 1.7\times 10^6$K and $n \sim 0.1 \cm3$ in a periodic 
box of 10 kpc and followed its evolution for up to 20 cooling time. 
We adopted idealized models in which a static, non-magnetized gas
cools radiatively with initial isothermal density perturbations.
A radiative cooling rate in ionization equilibrium for an optically
thin gas with the primordial composition was used. In addition, an
ad hot heating was included, which emulates feedbacks from stellar
winds, supernovae, turbulence and shocks in order to maintain
the thermal balance of background gas.
 
The main results can be summarized as follows:
1) Clouds form first on scales much smaller than the cooling length
as a result of non-linear behavior of TI.
2) Those small clouds grow through compression by background pressure
as well as gravitational infall, but eventually they merge by gravity to
become gravitationally bound objects.
3) The gravitationally bound clouds have acquired angular momentum through
merging as well as tidal torque. So they have high angular momentum with
the spin parameter of $\left<\lambda_s\right> \sim 0.3$ or so.
4) The spatial distribution of clouds depends on initial perturbations,
for instance, the slope of initial density power spectrum, but the
properties of individual clouds are not sensitive to that.

We note that the realistic picture of thermal-gravitational instability
that in protogalactic halos should be more complex than in the numerical
models considered here. Some of key aspects include: 
1) The gas in real protogalactic halos is likely 
in a chaotic state induced during the formation of halos themselves. 
The chaotic flow motions would have suppressed the early formation of clouds
via TI, but increased collisions of clouds once formed. 
2) Protogalactic halos have their own structures, but the effects of
those structures were ignored. For instance,
the tidal torque exerted by the halo and/or the rotation in disk
would have suppressed the formation of clouds.
3) There are emerging evidences that magnetic field existed even in
the early galaxies where the oldest stars formed
\citep[see, \eg][]{zwei02}. Then, undoubtedly the magnetic field should
have affected the formation and properties of clouds profoundly.
4) It is well known that when a hot gas cools from $T>10^6$ K,
it recombines out of ionization equilibrium because the cooling time
scale is shorter than the recombination time scale \citep{shkang87}.
However, details of the cooling such as non-equilibrium ionization
and \Htwo and metal cooling below $10^4$K could have only minor effects
on our main conclusions.

Although the numerical models are rather idealized to facilitate
simulations, our results should provide crude insights on the formation
of PGCCs in protogalactic halos.
The gravitationally bound clouds in our simulations have mass
$M_c \ga 10^7 \Msun$ and radius $R_c \simeq 150-200$ pc. If some of them
evolved into PGCCs and became globular clusters, they should have lost
$\sim 90 \%$ of their mass with $\sim 10 \%$ of star formation efficiency,
and at the same time they should have collapsed by a factor 10 or so.
But the further collapse would not have been straightforward because
of large angular momentum, unless their angular momentum was removed very
efficiently along with mass during the star formation phase. Such
removal of angular momentum may not be impossible, but following it is
beyond the scope of this numerical study. It should be studied with
simulations that have resolution high enough to follow the fragmentation
of clouds and the subsequent formation of stars inside PGCCs. However,
we make the following note of caution. The clouds in our simulations
continue to grow mostly through merging in late stage, and thus their
mass and angular momentum increase in time. So any simulations of
{\it isolated} clouds to study the ensuing evolution of PGCCs could be
misleading.

We should point that there is a caveat in our argument for the
large angular momentum of gravitationally bound clouds. The angular
momentum was acquired mostly through gravitational processes,
\ie tidal torque and merging. So if the clouds formed in an environment
where the gravitational processes are less important, they would have
acquired less angular momentum. For instance, we showed that the clouds
formed in a denser background have smaller angular momentum. Hence,
if protogalactic halos consisted of smaller halo-lets and clouds formed
in shocked regions after collisions of halo-lets, as suggested by \eg\ 
\citet{gunn80}, they could have smaller angular momentum.

\acknowledgments{
We thank the anonymous referee for constructive comments.
This work by CHB, HK and JK was supported by KOSEF through Astrophysical
Research Center for the Structure and Evolution of Cosmos (ARCSEC).
This work by DR was supported by Korea Research Foundation Grant
(KRF-2004-015-C00213). Numerical simulations were performed using
``Linux Cluster for Astronomical Calculation'' of KASI-ARCSEC.}

\appendix

\section{Corrected Potential}

In our simulations, the gravitational potential for the gas-dynamical
equation in \S 2.2 was calculated by using the FFT method,
$\Phi_{\rm FFT}(\vec r)$. Then the gravitational force could be
correctly calculated by differentiating this potential on the grid.
However the use of $\Phi_{\rm FFT}(\vec r)$ in calculating the
gravitational energy of clouds, $E_G$, in \S 3.3 ends up a large error,
because $\Phi_{\rm FFT}(\vec r)$ is undetermined by an {\it integral
constant}. In principle, the gravitational potential of isolated clouds
can be precisely calculated by the direct double integration over cloud
volume. However, the computational cost of this method is prohibitively
expensive, especially for gravitationally bound clouds in the $1024^3$
simulation, since they occupy typically $\sim 10^4$ or so grid zones.
One the other hand, the direct integration does not take account of
contributions from the rest of mass in the simulation box as well as
the periodic mass distribution. But we found that those contributions
are small, especially for massive, gravitationally bound clouds.

As an effort to estimate the gravitational energy of clouds more
accurately, we devised a method which calculates and uses the corrected
potential as follows:\\
1) The position, $\vec r_{\rm min}$, where $\Phi_{\rm FFT}(\vec r)$
has the minimum value, is found for each cloud, and then the potential
at the position is calculated by the direct integration,
\begin{equation}
\Phi_{\rm direct}(\vec r_{\rm min}) = - G \int {{\rho({\vec {r'}})} \over
{|{\vec {r'}}-\vec r_{\rm min}|}} dV'.
\end{equation}
2) The difference between $\Phi_{\rm direct}$ and
$\Phi_{\rm FFT}$ at $\vec r_{\rm min}$ is calculated for each cloud,
\begin{equation}
C = \Phi_{\rm direct}(\vec r_{\rm min}) - \Phi_{\rm FFT}(\vec r_{\rm min}).
\end{equation}
3) Then the corrected potential for each cloud is calculated by 
\begin{equation}
\Phi_{\rm corr}(\vec r) = \Phi_{\rm FFT}(\vec r) + C.
\end{equation}
4) Finally the gravitation energy of each cloud is calculated as
\begin{equation}
E_G = \int { 1 \over 2 } \rho(\vec r) \Phi_{\rm corr}(\vec r) dV.
\end{equation}

Figure 15 demonstrates the motivation of our effort. Here $E_{\rm N^2}$ is
the gravitational energy of clouds calculated by the direct double
integration, while $E_{\rm FFT}$ and $E_{\rm corr}$ were calculated using
$\Phi_{\rm FFT}$ and $\Phi_{\rm corr}$, respectively. The errors for the
S0256 model are shown, where the double integration could be done with
a reasonable computation time. $E_{\rm corr}$ tends to agree with
$E_{\rm N^2}$ better than $E_{\rm FFT}$. With $E_{\rm corr}$ the error
is within $\sim 20 \%$ or so for a substantial fraction of clouds.
However, the estimation of gravitational energy could be easily off by
a factor two or even larger with $E_{\rm FFT}$. We used $E_{\rm corr}$
for the gravitational energy in \S 3.3.

\clearpage

\begin{deluxetable}{ccccccc}
\tablecaption{Model Parameters for Simulations\tablenotemark{a}}
\tablehead{\colhead{Model} & \colhead{No. of grid zones}
& \colhead{$t_{\rm end}$ ($\tcl$)\tablenotemark{b}}
& \colhead{$t_{\rm end}$ ($t_{\rm grav}$)\tablenotemark{c}}
& \colhead{$T_{\rm min}$ (K)}
& \colhead{$P_k$} & \colhead{$n_h$ (${\rm cm}^{-3}$)\tablenotemark{d}} }
\startdata
S1024 & $1024^3$ & 16 & 2.29 & $10^4$ & const & 0.1 \\
S0512 & $512^3$  & 20 & 2.86 & $10^4$ & const & 0.1 \\
S0256 & $256^3$  & 20 & 2.86 & $10^4$ & const & 0.1 \\
K0512 & $512^3$  & 20 & 2.86 & $10^4$ & $\propto k^{-{5 \over 3}}$ & 0.1 \\
R0512 & $512^3$  & 20 & 2.86 & $10^4$ & $\propto k^2$ & 0.1 \\
C0512 & $512^3$  & 20 & 2.86 & $10^2$ & const & 0.1 \\
D0512 & $512^3$  & 20 & 1.65 & $10^4$ & const & 0.3 \\
\enddata
\tablenotetext{a}{Simulation box size $L = 10$ kpc in all models.}
\tablenotetext{b}{$\tcl=2\times 10^7$ yrs in the models with
$n_h = 0.1 \cm3$, and $\tcl=6.7\times 10^6$ yrs in the model with
$n_h = 0.3 \cm3$.}
\tablenotetext{c}{$t_{\rm grav} = 1.4 \times 10^8$ yrs in the models
with $n_h = 0.1 \cm3$, and $t_{\rm grav} = 8.1 \times 10^7$ yrs in
the model with $n_h = 0.3 \cm3$.}
\tablenotetext{d}{$\rho_h = (2.34\times 10^{-24}{\rm g})n_h$ with
$n({\rm He})/n({\rm H}) = 0.1.$}
\end{deluxetable}

\clearpage

\begin{figure}
%\plotone{f1.eps}
\vspace{-8cm}\hspace{-7cm}\epsfxsize=30cm\epsfbox{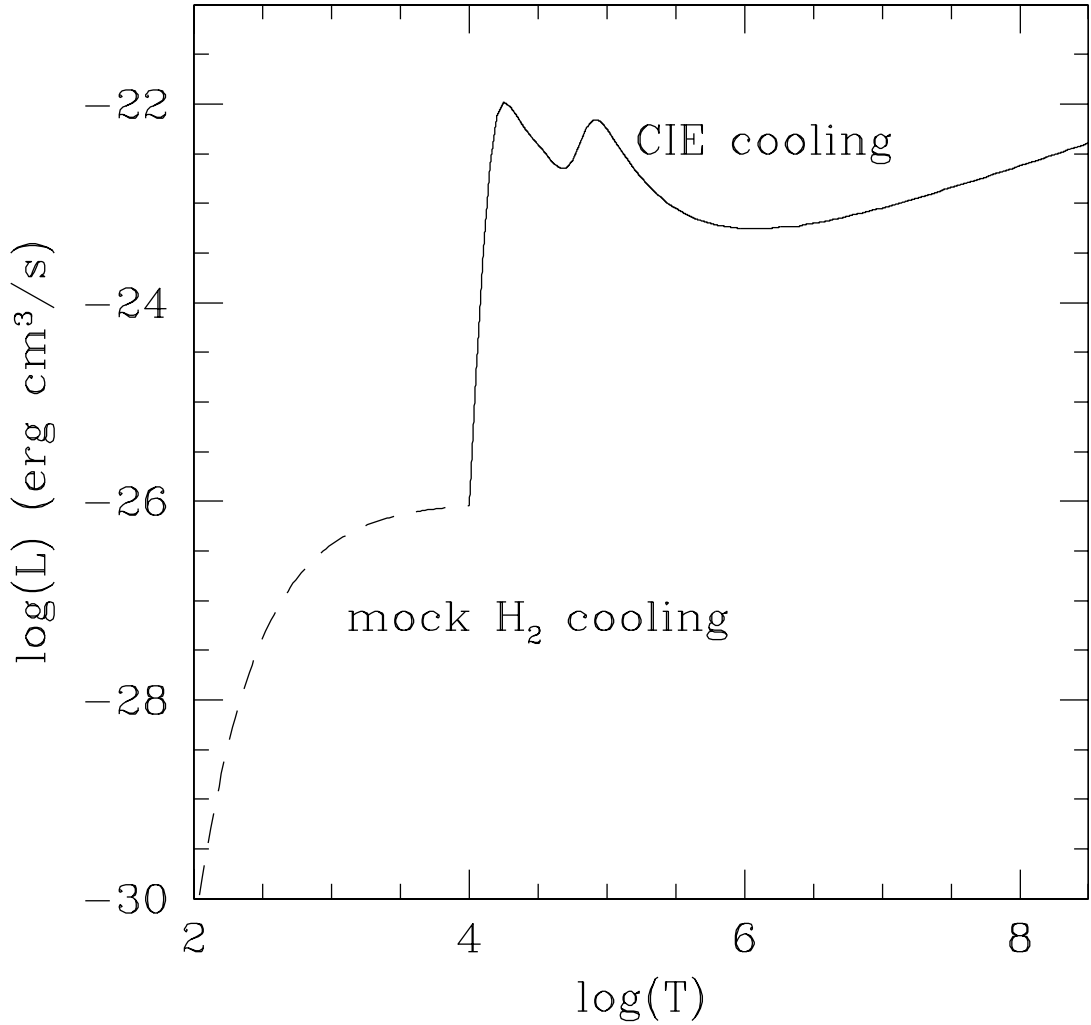}\vspace{-8cm}
\figcaption
{Cooling function, $L(T) \equiv \Lambda/n_H^2$, of the collisional
ionization equilibrium (CIE) cooling for a gas with zero-metalicity
(solid line). The dotted line is for the mock \Htwo cooling adopted
in Model C.}
\end{figure}

\begin{figure}
%\plotone{f2.eps}
\vspace{-10cm}\hspace{-9cm}\epsfxsize=34cm\epsfbox{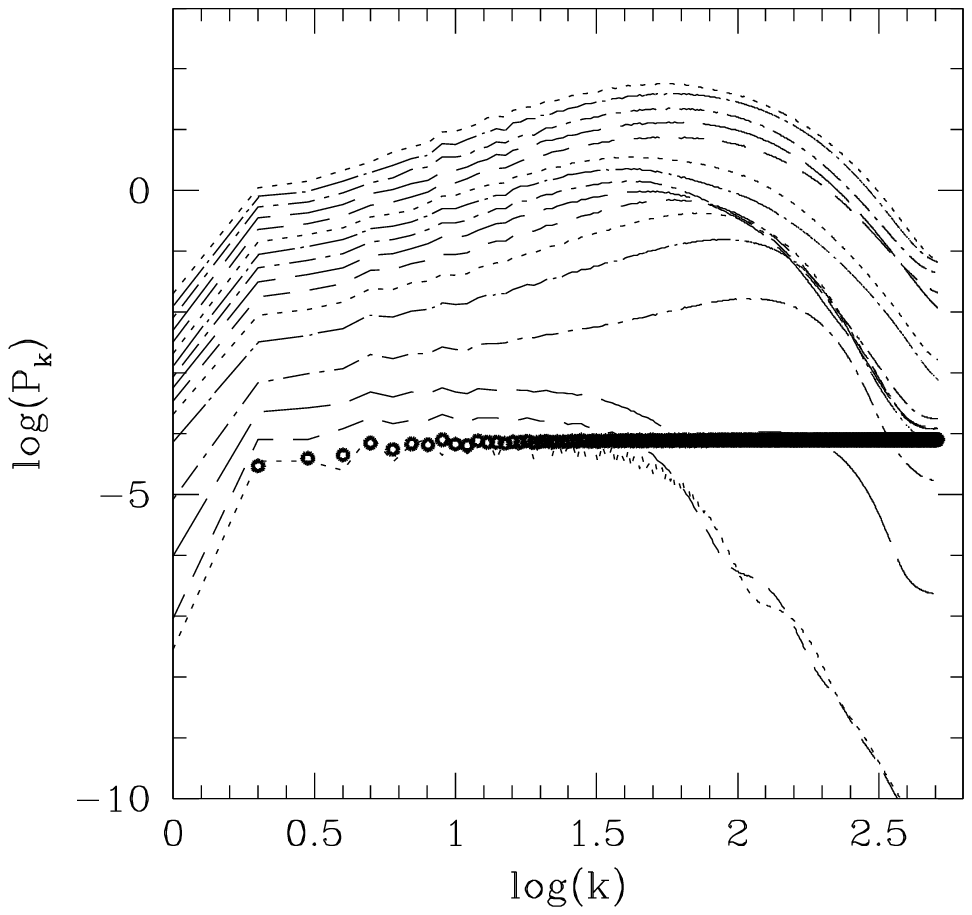}\vspace{-10cm}
\figcaption
{Evolution of the density power spectrum in the S1024 model. Dots
represent the initial power spectrum at $t = 0$ and lines are the power
spectrum at $\tcl$, $2~\tcl$, $3~\tcl$, $\ldots$, $16~\tcl$.}
\end{figure}

\begin{figure}
\plotone{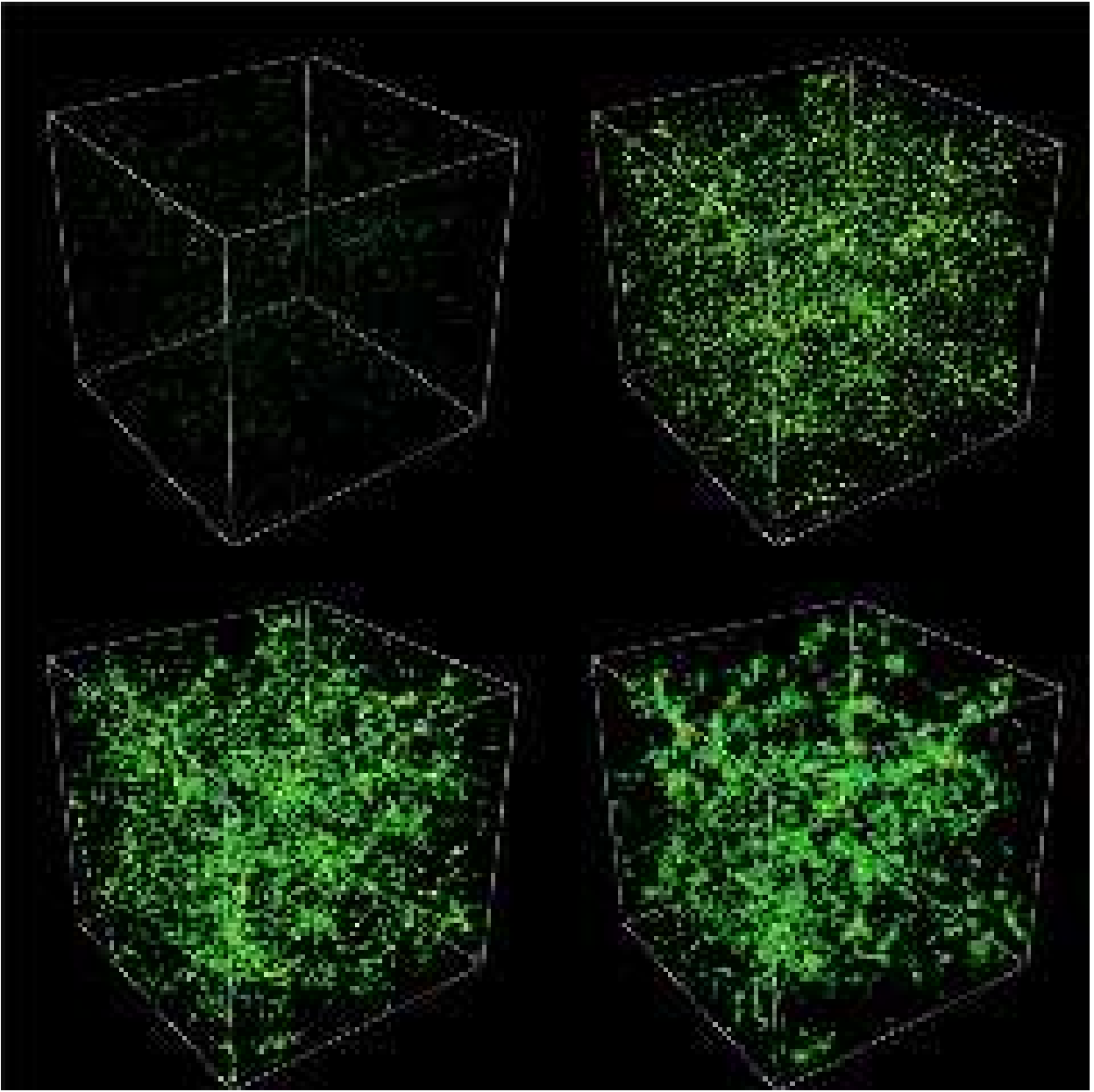}
\figcaption
{Iso-density surfaces inside the full simulation box of size $10 \kpc$ at
$4~\tcl$ ({\it top-left}), $8~\tcl$ ({\it top-right}), $12~\tcl$
({\it bottom-left}), and $16~\tcl$ ({\it bottom-right}) in the S1024 model.
Green surfaces corresponds to $10\rho_0$, yellow surfaces corresponds to
$10^2\rho_0$ and red surfaces corresponds to $10^3 \rho_0$. Here $\rho_0$
is the mean initial density.}
\end{figure}

\begin{figure}
%\plotone{f4.eps}
\vspace{0cm}\hspace{0.5cm}\epsfxsize=15cm\epsfbox{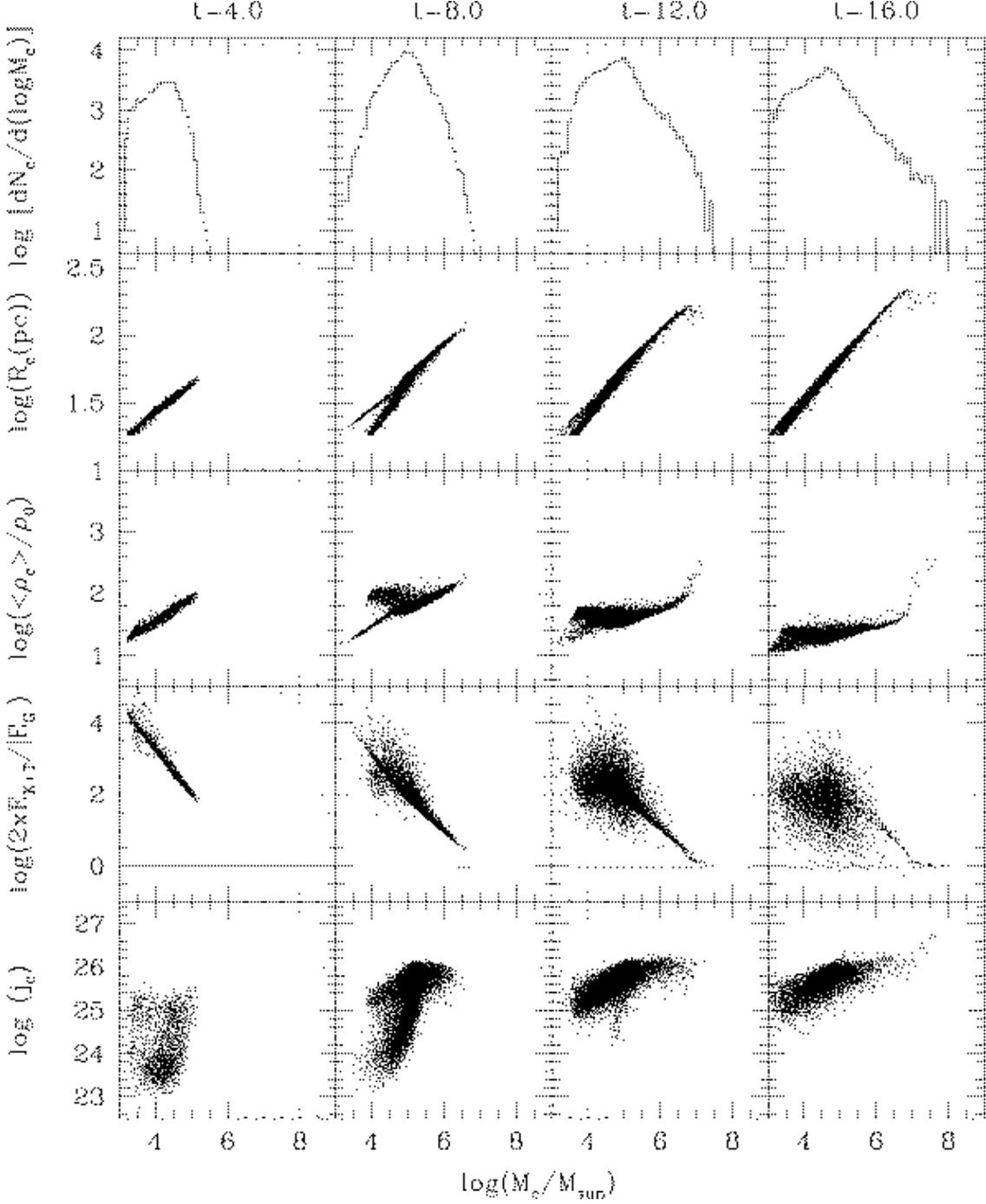}\vspace{0cm}
\figcaption
{Differential number of clouds, $dN_c/d(\log M_c)$, effective radius, $R_c$,
mean density, $\left<\rho\right>_c$, energy ratio, $2 (E_T + E_K) / |E_G|$,
and specific angular momentum, $j_c$, as a function of cloud mass, $M_c$,
at four different times in the S1024 model. Here, $j_c$ is in the cgs units}
\end{figure}

\begin{figure}
%\plotone{f5.eps}
\vspace{-9cm}\hspace{-8.5cm}\epsfxsize=33cm\epsfbox{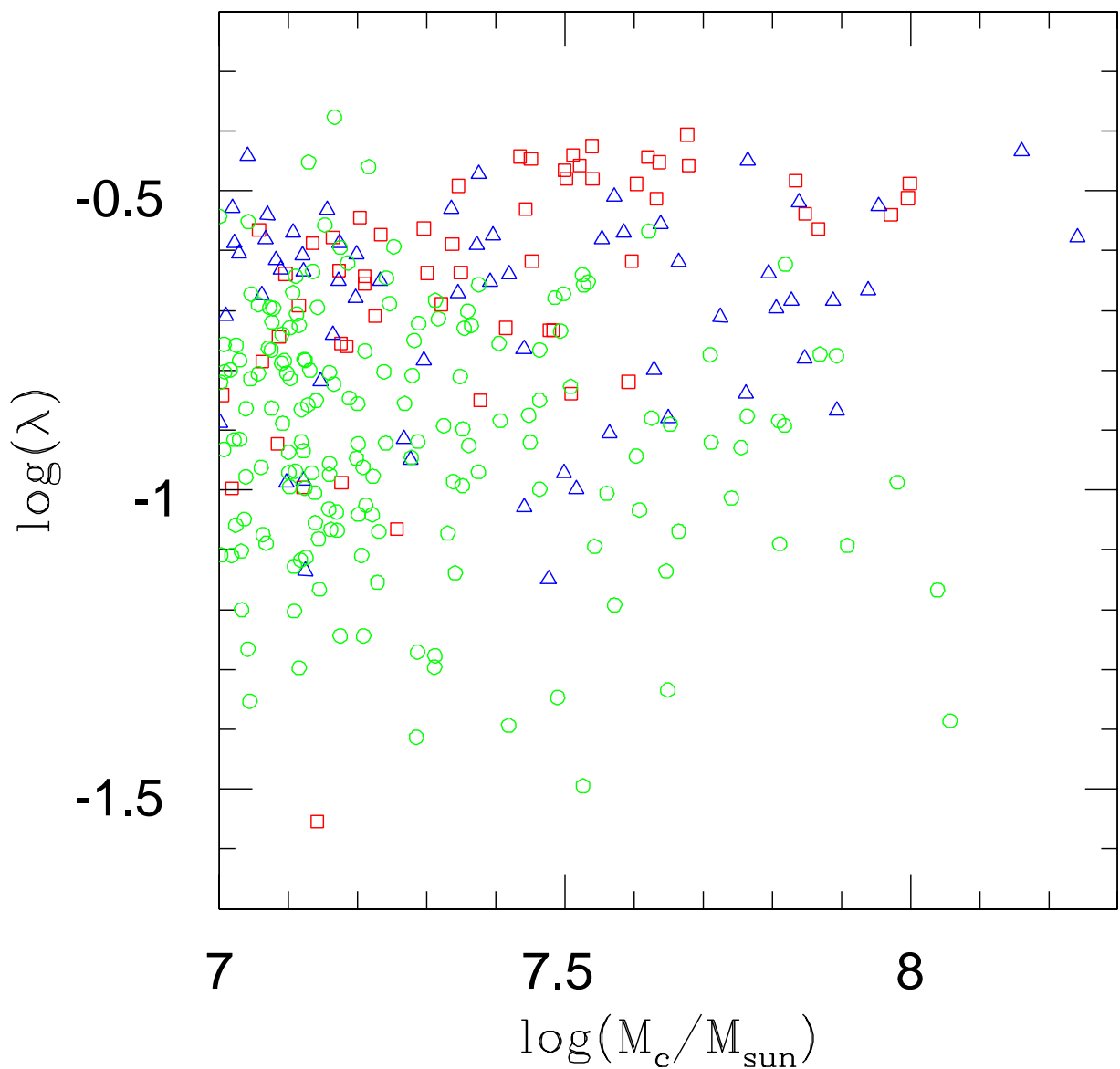}\vspace{-9cm}
\figcaption
{Spin parameter, $\lambda_s = {{J|E_{G}^{1/2}|} / {GM^{5/2}}}$, as a
function of cloud mass, $M_c$, for clouds with $M_c \geq 10^7 \Msun$.
Red squares are for the S1024 model, blue triangles are for the S0512
model, and green circles are for the D0512 model at $16~\tcl$.}
\end{figure}

\begin{figure}
% \plotone{f6.eps}
\vspace{-3cm}\hspace{-2cm}\epsfxsize=22cm\epsfbox{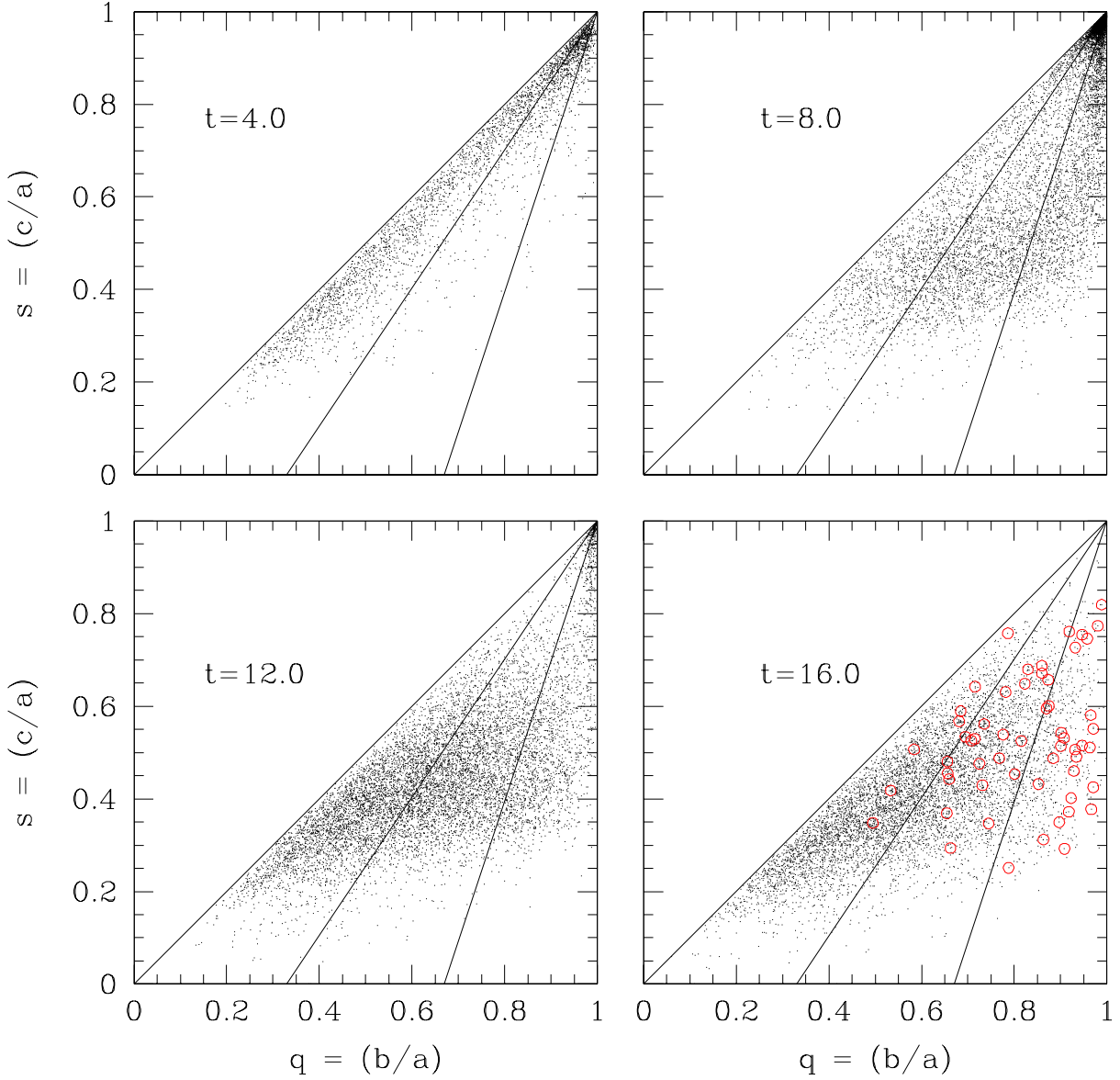}\vspace{-2cm}
\figcaption
{Shape parameters, $q=b/a$ and $s=c/a$, at four different times in
the S1024 model. The principal axes of clouds are defined such that
$a \geq b \geq c$. The domain is divided into three regions containing
the clouds of prolate shape ($s \sim q < 1$, {\it left}), and the clouds
of oblate shape ($s < 1$ and $q \sim 1$, {\it right}) and the clouds of
triaxial shape ($s < q < 1$, {\it middle}). Red circles represent the
clouds with $M_c \geq 10^7 \Msun$.}
\end{figure}

\begin{figure}
% \plotone{f7.eps}
\vspace{-8cm}\hspace{-3.5cm}\epsfxsize=24cm\epsfbox{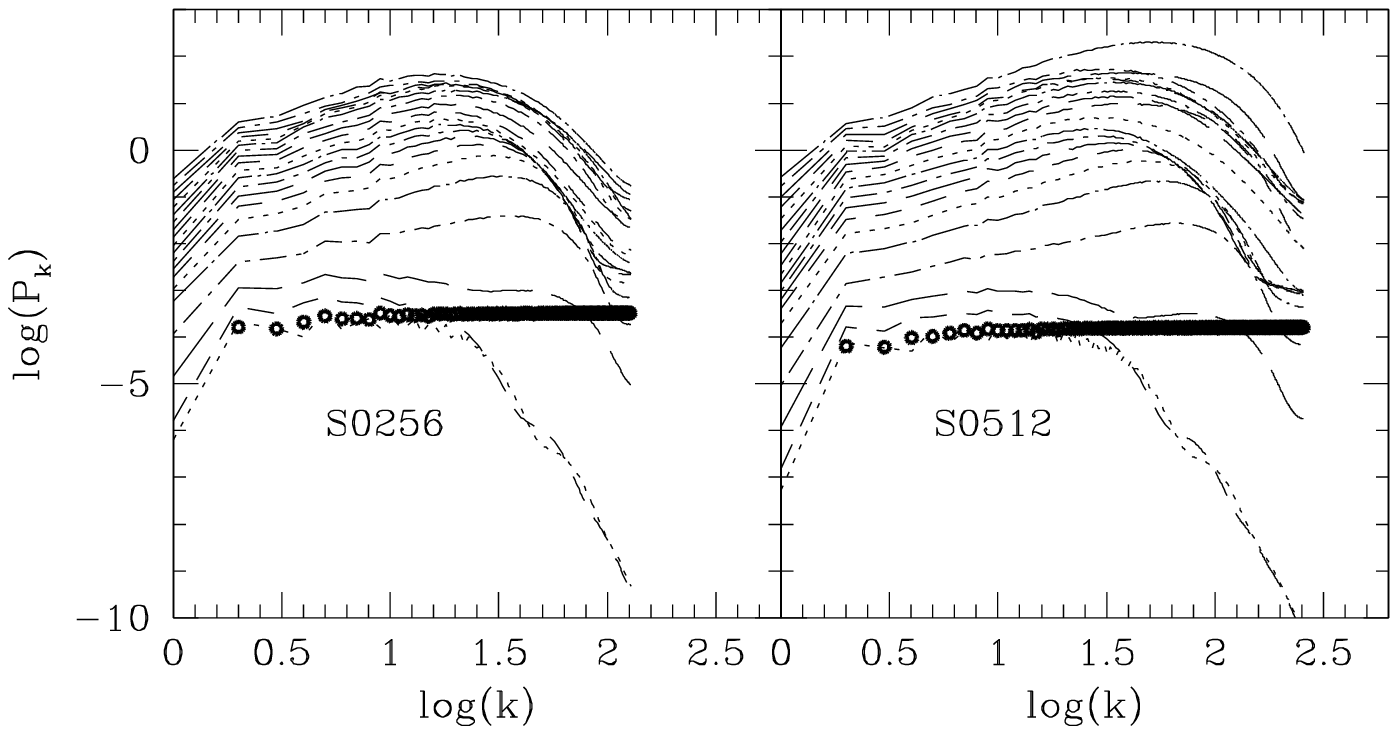}\vspace{-7cm}
\figcaption
{Evolution of the density power spectrum in the S0256 and S0512 models.
Dots represent the initial power spectrum at $t = 0$ and lines are the
power spectrum at $\tcl$, $2~\tcl$, $3~\tcl$, $\ldots$, $20~\tcl$.}
\end{figure}

\begin{figure}
% \plotone{f8.eps}
\vspace{-1cm}\hspace{-1.5cm}\epsfxsize=20cm\epsfbox{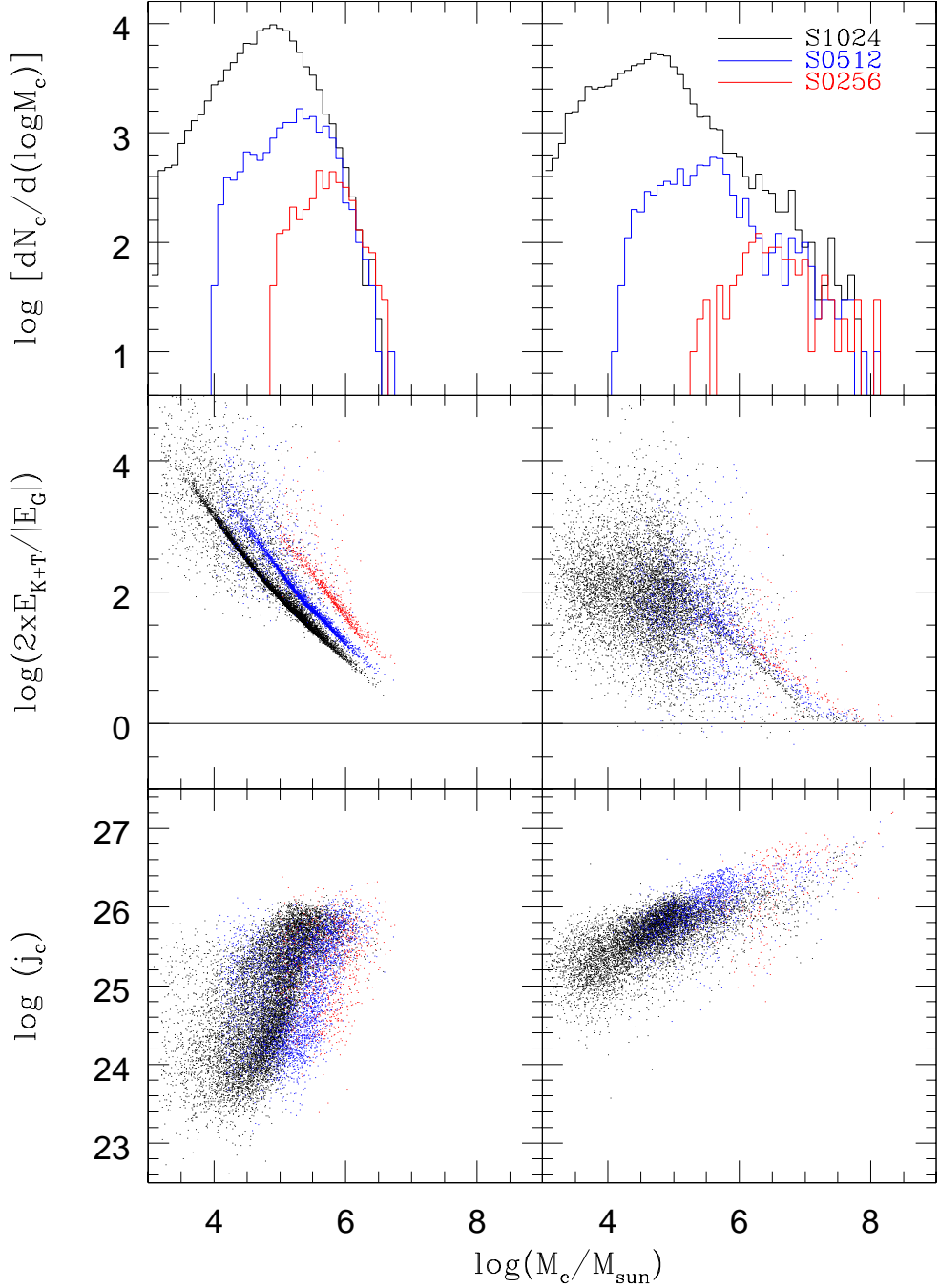}\vspace{-1cm}
\figcaption
{Differential number of clouds, $dN_c/d(\log M_c)$, energy ratio,
$2 (E_T + E_K) / |E_G|$, and specific angular momentum, $j_c$, as a
function of cloud mass, $M_c$, in the three models with different
resolution. {\it Left} panels show the quantities at $5~\tcl$,
$6~\tcl$, and $7~\tcl$ for the S0256, S0512, and S1024 models,
respectively, and {\it right} panels show the quantities at
$15~\tcl$ for all three models. Here, $j_c$ is in the cgs units.}
\end{figure}

\begin{figure}
% \plotone{f9.eps}
\vspace{-8cm}\hspace{-3.5cm}\epsfxsize=24cm\epsfbox{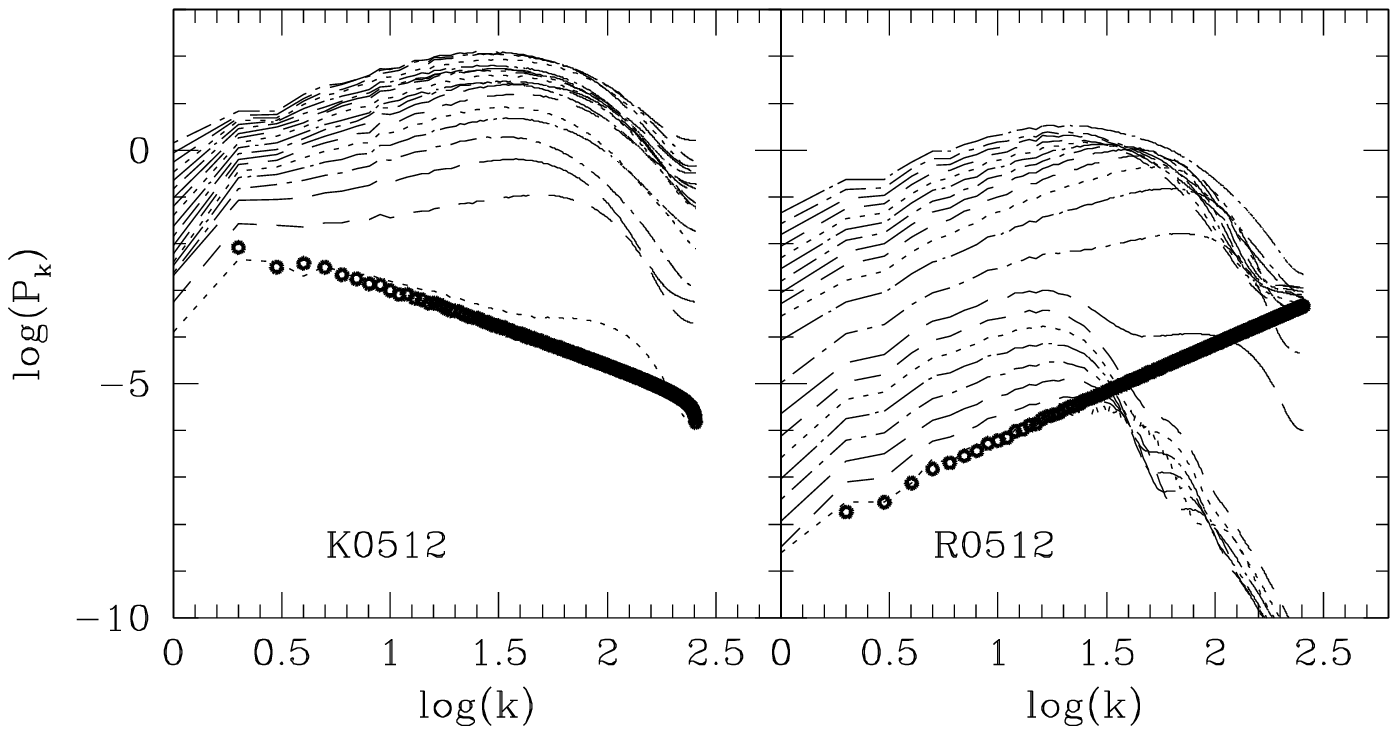}\vspace{-7cm}
\figcaption
{Evolution of the density power spectrum in the K0512 and R0512 models.
Dots represent the initial power spectrum at $t = 0$ and lines are the
power spectrum at $\tcl$, $2~\tcl$, $3~\tcl$, $\ldots$, $20~\tcl$.}
\end{figure}

\begin{figure}
% \plotone{f10.eps}
\vspace{-9cm}\hspace{-8.5cm}\epsfxsize=33cm\epsfbox{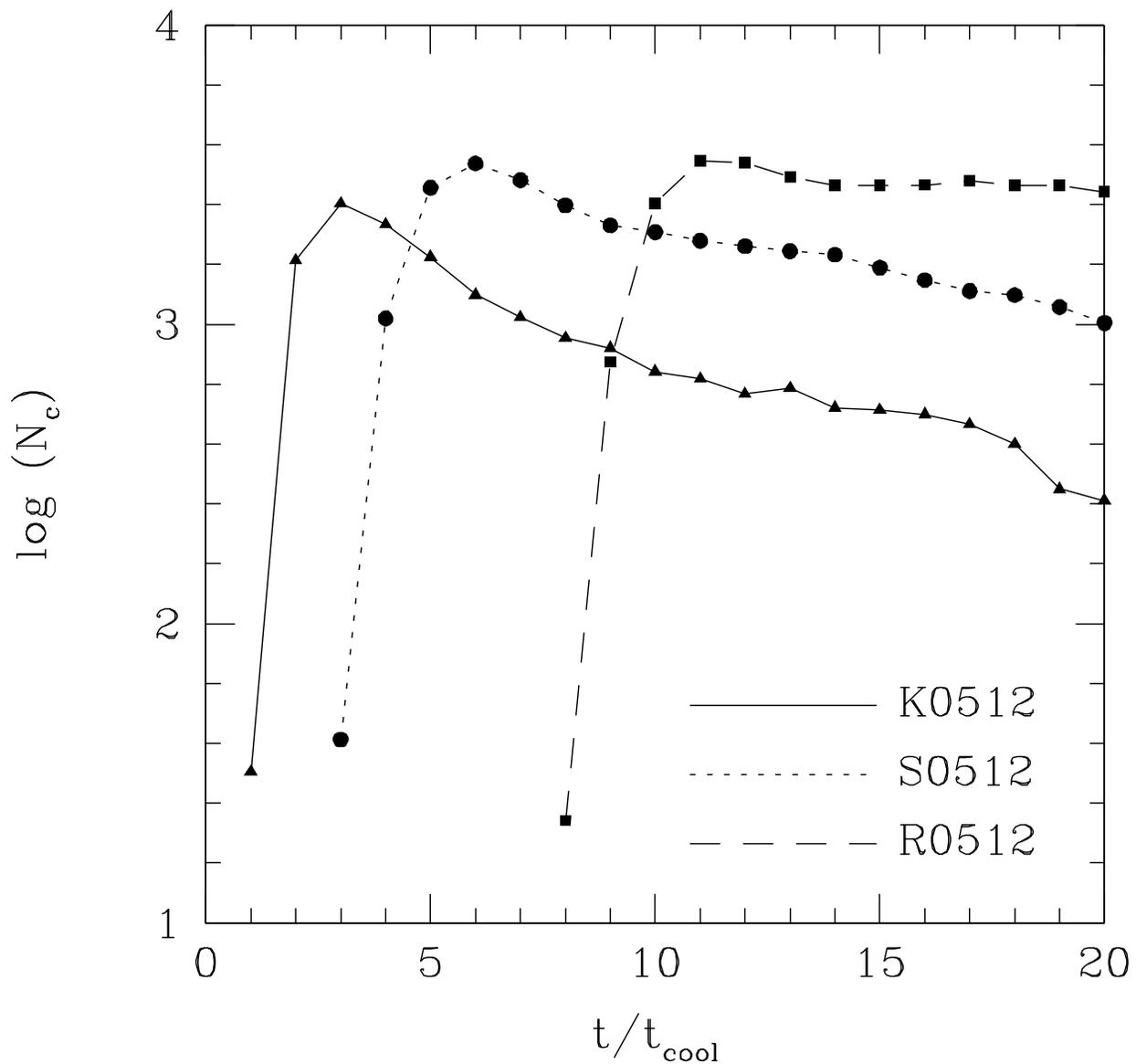}\vspace{-9cm}
\figcaption
{Time evolution of the number of clouds, $N_c$, in the three models with
different initial density power spectrum.}
\end{figure}

\begin{figure}
\plotone{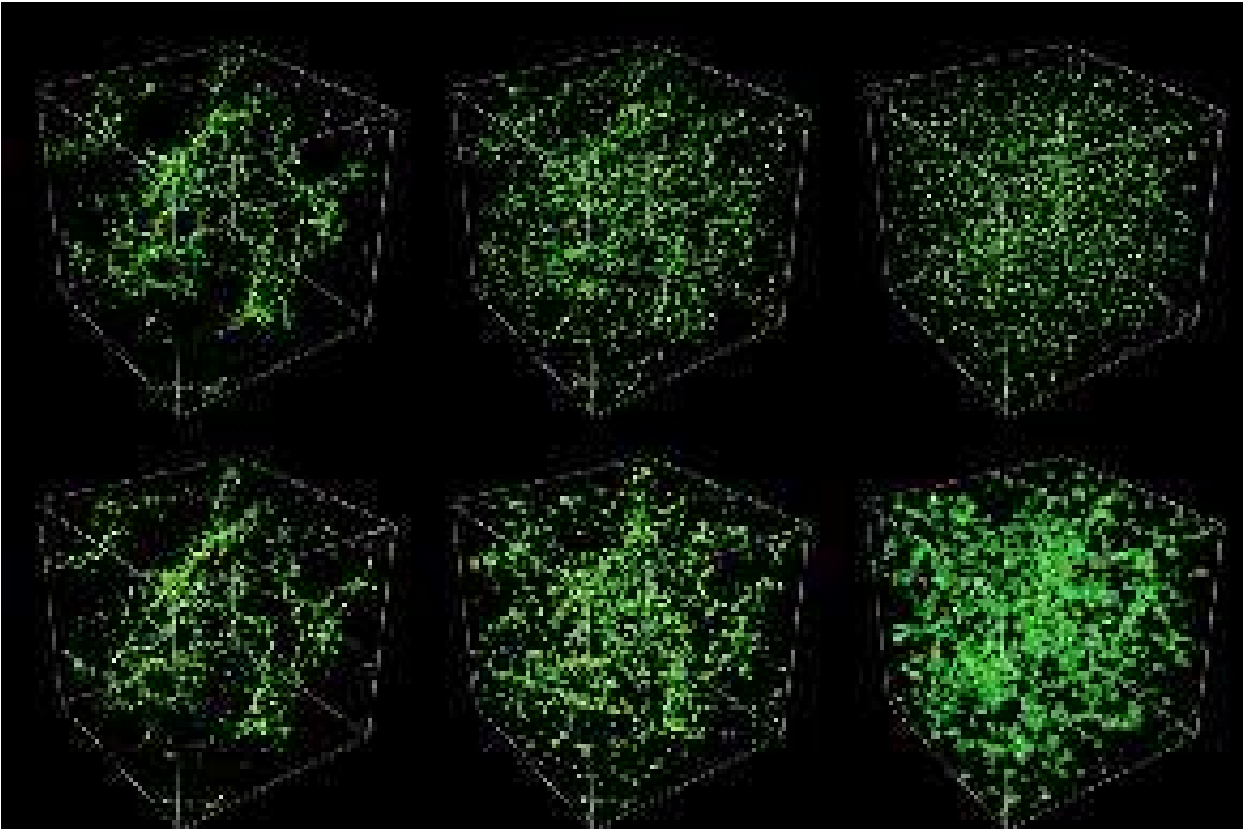}
\figcaption
{Iso-density surfaces inside the full simulation box of size $10 \kpc$
in the K0512 model ({\it left} images), in the S0512 model ({\it middle}
images), and in the  R0512 model ({\it right} images). {\it Top} images
are at $3~\tcl$ in the K0512 model, at $6~\tcl$ in the S0512 model,
and at $11~\tcl$ in R0512 model. {\it Lower} images are at $4~\tcl$
in the K0512 model, at $10~\tcl$ in the S0512 model, and at $19~\tcl$
in the R0512 model. Green surfaces corresponds to $10\rho_0$, yellow
surfaces corresponds to $10^2\rho_0$ and red surfaces corresponds to
$10^3 \rho_0$, same as in Figure 3.}
\end{figure}

\begin{figure}
% \plotone{f12.eps}
\vspace{-1cm}\hspace{-1.5cm}\epsfxsize=20cm\epsfbox{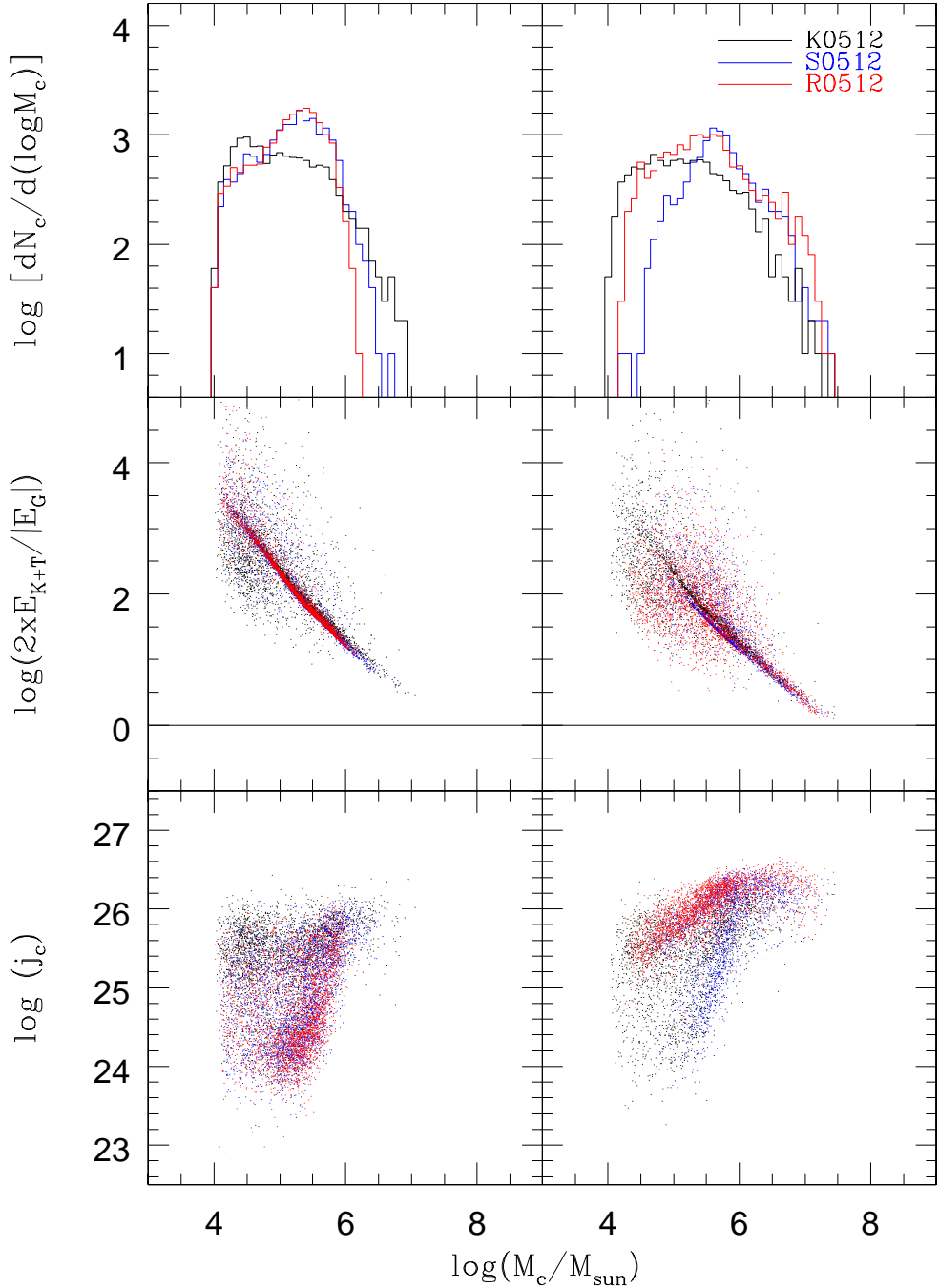}\vspace{-1cm}
\figcaption
{Differential number of clouds, $dN_c/d(\log M_c)$, energy ratio,
$2 (E_T + E_K) / |E_G|$, and specific angular momentum, $j_c$, as a
function of cloud mass, $M_c$, in the three models with different
initial density power spectrum. {\it Left} panels are at $3~\tcl$,
$6~\tcl$, and $11~\tcl$ for the K0512, S0512, and R0512 models,
respectively, and {\it right} panels are at $4~\tcl$, $10~\tcl$,
and $19~\tcl$ for the K0512, S0512, and R0512 models, respectively.
Here, $j_c$ is in the cgs units.}
\end{figure}

\begin{figure}
% \plotone{f13.eps}
\vspace{-1cm}\hspace{-1.5cm}\epsfxsize=20cm\epsfbox{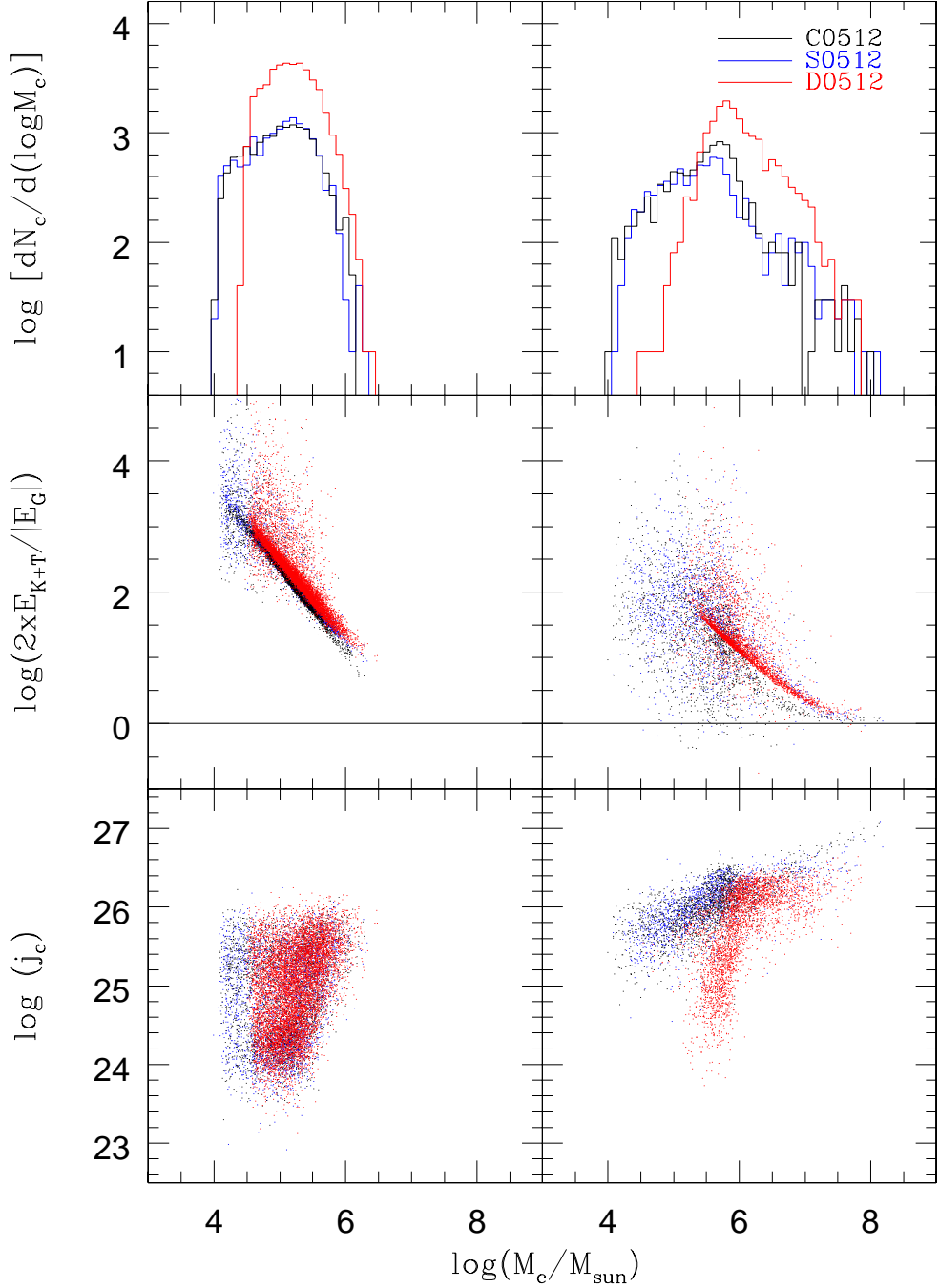}\vspace{-1cm}
\figcaption
{Differential number of clouds, $dN_c/d(\log M_c)$, energy ratio,
$2 (E_T + E_K) / |E_G|$, and specific angular momentum, $j_c$, as a
function of cloud mass, $M_c$, in the C0512 and D0512 models as well as
in the S0512 model for comparison. {\it Left} panels are at $5~\tcl$,
and {\it right} panels are at $15~\tcl$, respectively, for all three
models. Here, $j_c$ is in the cgs units.}
\end{figure}

\begin{figure}
% \plotone{f14.eps}
\vspace{-7cm}\hspace{-2.5cm}\epsfxsize=21.5cm\epsfbox{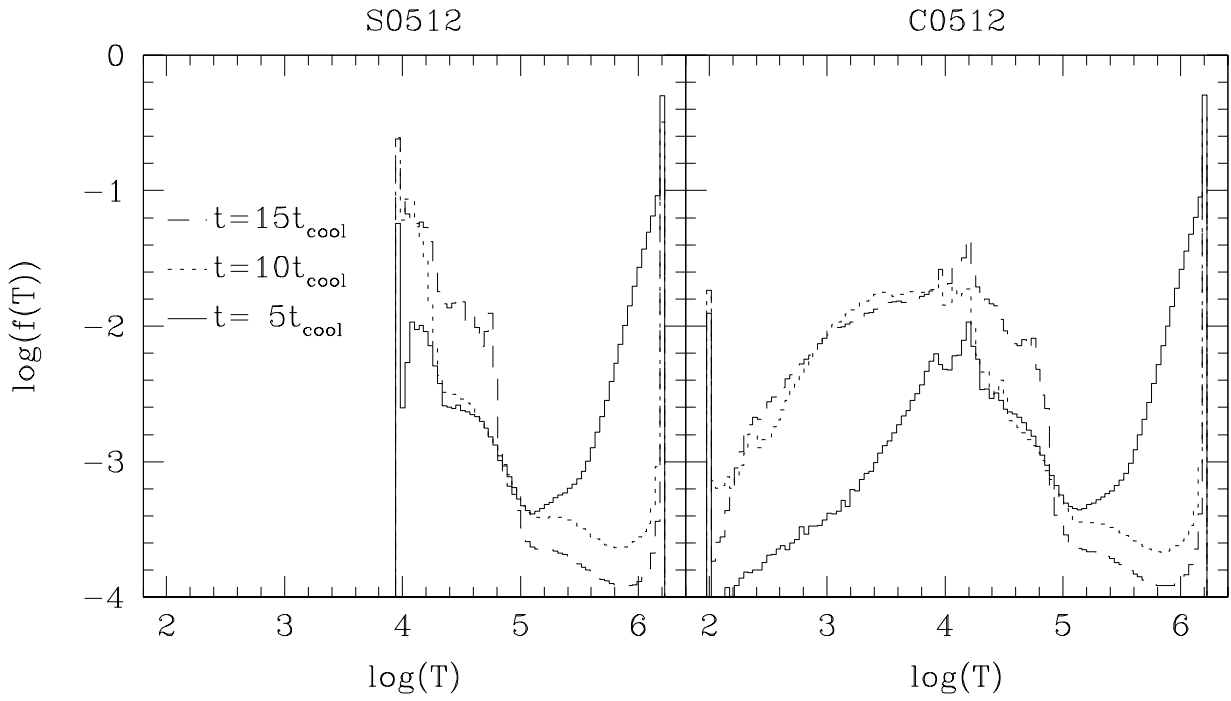}\vspace{-6cm}
\figcaption
{Mass fraction, $f( T)$, of gas as a function of temperature at three
different times in the S0512 and C0512 models.}
\end{figure}

\begin{figure}
% \plotone{f15.eps}
\vspace{-3cm}\hspace{-2cm}\epsfxsize=22cm\epsfbox{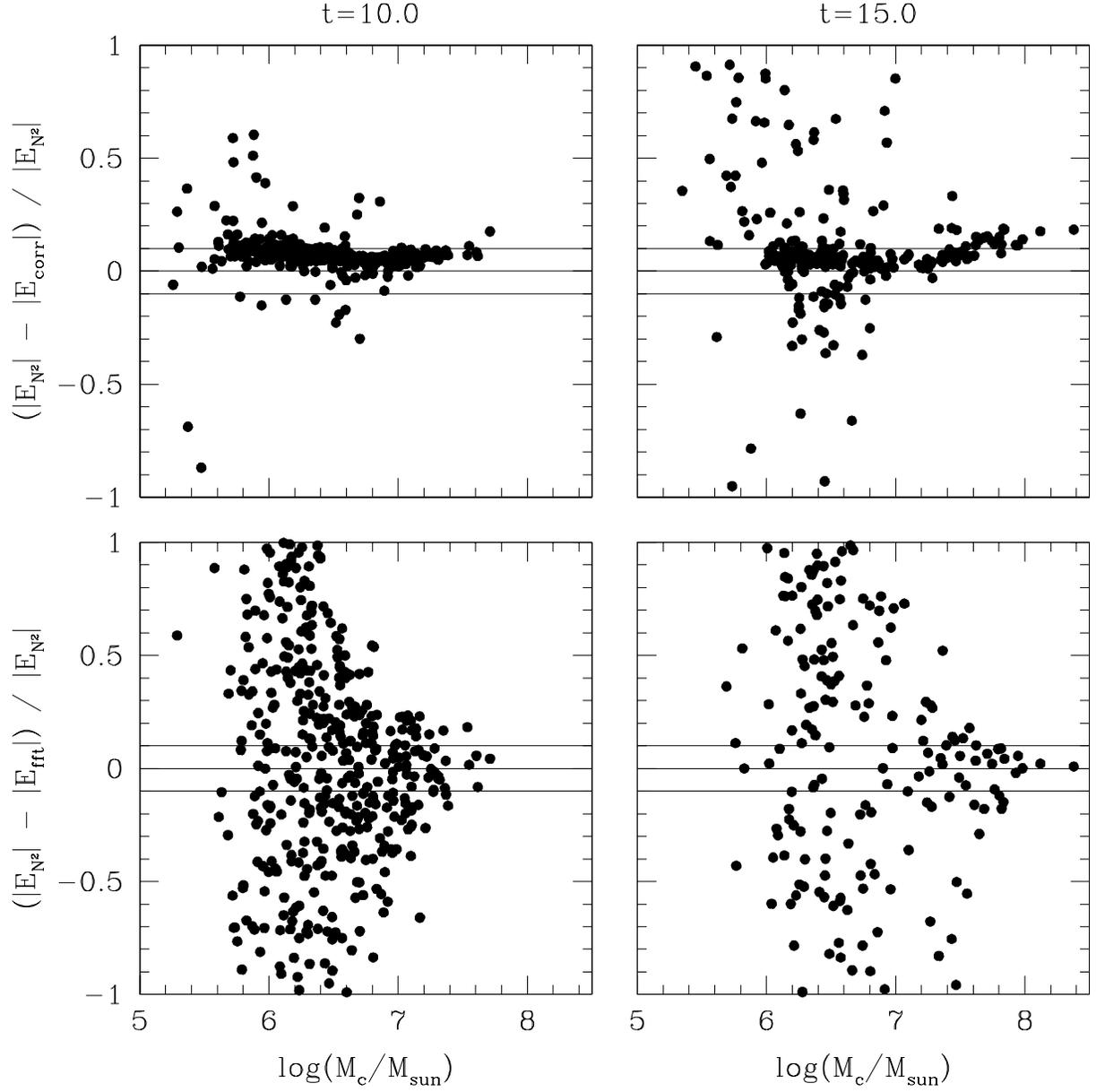}\vspace{-2cm}
\figcaption
{Error in the estimation of gravitational energy by the FFT potential
({\it bottom} panels) and the corrected potential ({\it top} panels) 
at $10~\tcl$ ({\it left} panels) and $15~\tcl$ ({\it right} panels)
in the S0256 model. See the Appendix A for details.}
\end{figure}

\end{document}